\documentclass[journal]{IEEEtran}

\usepackage{amsmath,amssymb,amsfonts,amsthm,mathrsfs}
\usepackage{algorithm,algorithmic}
\usepackage{graphicx}
\usepackage{textcomp}
\usepackage{color,xcolor}
\usepackage{bm}
\usepackage{multirow,booktabs}
\usepackage[normalem]{ulem}
\usepackage{tabularx}
\usepackage{xpatch}
\usepackage{soul}
\usepackage{ulem}
\usepackage{verbatim}
\usepackage{subcaption}
\usepackage{enumitem}
\usepackage{hyperref}
\usepackage[acronym]{glossaries}
\usepackage[linewidth=0.8pt]{mdframed}
\usepackage{cite}

\usepackage{array}

\hypersetup{
	colorlinks=true,  
	linkcolor=red,    
	citecolor=blue,   
	urlcolor=black   
}

\theoremstyle{theorem,lemma,remark,proposition}

\ifCLASSINFOpdf
\else
\fi

\begin{document}

\title{
Rydberg Atomic Quantum Receivers for OFDM Systems:~An Atomic-Comb Design
}

\author{Yuanbin~Chen,~Chong~Meng~Samson~See,~\IEEEmembership{Member,~IEEE,}~and~Chau~Yuen,~\IEEEmembership{Fellow,~IEEE}


\thanks{
	Yuanbin Chen and Chau Yuen are with the School of Electrical and Electronics Engineering, Nanyang Technological University, Singapore 639798 (emails: yuanbin.chen@ntu.edu.sg; chau.yuen@ntu.edu.sg).

	Chong Meng Samson See is with DSO National Laboratories, Singapore 118225 (e-mail: schongme@dso.org.sg).


}
	
	
}

\maketitle

\begin{abstract}
Rydberg atomic quantum receivers (RAQR) exhibit exceptional sensitivity to radio-frequency (RF) signals, positioning them as promising candidates for next-generation wireless systems. Extending their capabilities to orthogonal frequency-division multiplexing (OFDM), however, requires atomic responses that go beyond broadband reception and are tailored to the structured interference of multicarrier signals. To bridge this gap, we develop an atomic-comb Rydberg receiver based on six-wave mixing (SWM) transduction. In particular, injecting a multi-tone local oscillator (LO) into the RAQR generates frequency-shifted replicas of the intrinsic SWM response. By subtly tuning the amplitude and phase of each LO tone, constructive and destructive interference can be selectively induced to preserve desired response components while attenuating harmful frequency offsets. Furthermore, we harness the response-reshaping capability of the proposed atomic comb to proactively mitigate OFDM interference. More explicitly, by exploiting the discrete difference-frequency structure of both carrier-frequency-offset (CFO)-induced and phase-noise (PN)-induced inter-carrier interference (ICI), we optimize the atomic-comb response to suppress dominant ICI components directly. We further reveal an intrinsic coupling between spectral shaping and the time-domain effective channel memory, motivating a joint optimization of both ICI and inter-symbol interference (ISI), which simultaneously suppresses inter-carrier leakage and post-cyclic-prefix (CP) tail energy.
Extensive simulation results validate our atomic-comb design, unveiling that the proposed atomic comb can effectively reshape the receiver response over the OFDM, while mitigating ICI, shortening the effective channel memory, and substantially lowering CP overhead. 

\end{abstract}


%
%
%
%
%
%
%
%
%
%
%

%
\IEEEpeerreviewmaketitle

\section{Introduction}

The remarkable progress of quantum information technologies is redefining the role of quantum systems in information processing, sensing, metrology, and communications~\cite{QS-101}. 
At the heart of this expansion lies quantum sensing, which capitalizes on the exceptional responsiveness of well-controlled quantum states to external physical fields, providing fundamentally new mechanisms for electromagnetic (EM)-field detection and signal reception~\cite{QS-31}. In this context, highly excited Rydberg atoms have emerged as a particularly promising medium for radio-frequency (RF) sensing and wireless communications, benefiting from their giant electric dipole moments and exceptionally strong coupling to RF fields~\cite{terry-WCM,terry-trans,cui-CM,chen-Rydberg1}. By transducing RF-induced perturbations of atomic energy levels and quantum states into optically measurable signatures, Rydberg atoms establish a direct atomic-scale RF-to-optical interface that departs fundamentally from the conventional antenna-amplifier-mixer architecture. These unique characteristics have fueled growing interest in Rydberg atomic quantum receivers (RAQR) as quantum-enabled radio front ends, with the significant potential to reshape the design of future wireless communication and EM sensing systems~\cite{QS-3,peng-arxiv,peng-tvt}.

RAQR have evolved rapidly across several mutually reinforcing dimensions, spanning sensitivity, metrological traceability, spectral coverage, and communications. A major milestone is the introduction of atomic superheterodyne detection, which improves the experimentally demonstrated electric-field sensitivity to $55~\text{nV}/\text{cm}/\sqrt{\text{Hz}}$ while providing access to both the phase and frequency of incident RF signals~\cite{QS-16,QS-28}. More recently, quantum-enhanced electrometry employing cold Rydberg gases has pushed the equivalent sensitivity even further, reaching $1.6(5)~\text{nV}/\text{cm}/\sqrt{\text{Hz}}$~\cite{Wang2026Quantum-NC}. More fundamentally, the quantum projected noise floor of RAQR can approach the standard quantum limit (SQL), which is on the order of $10^{-2}~\text{nV}/\text{cm}/\sqrt{\text{Hz}}$ under favorable atomic-number and coherence conditions~\cite{QS-31,QS-19,QS-34}.
Equally important, the atomic nature of the sensing medium provides an intrinsic reference for self-calibrated and International system of units
(SI)-traceable electric-field measurements, with foundational experiments demonstrating operation over carrier frequencies ranging from approximately $1$ to $500~\text{GHz}$~\cite{QS-10-TAP}.
Beyond broad carrier-frequency accessibility, substantial progress has also been made toward frequency-agile and wide-instantaneous-bandwidth reception. In particular, recent microwave (MW)-to-optical-conversion experiments have pushed the instantaneous bandwidth of RAQR into the tens-of-megahertz regime while retaining high sensitivity through sideband engineering and six-wave mixing (SWM) mechanism~\cite{ref_1_yang2024,ref_2_PRA,QS-30,chen-Rydberg3,jian_xiao}.

Beyond single-carrier operation, recent studies have extended RAQR toward multifrequency, multiband, and multicarrier reception~\cite{MC-1,MC-2,MC-4,MC-6,MC-3,MC-5}. In particular, representative advancements include the simultaneous demodulation of multiple widely separated RF carriers using a single Rydberg atomic sensor~\cite{MC-6} and high-sensitivity phase-modulated frequency-division-multiplexing (FDM) communication~\cite{MC-3}. More recently, optical-frequency-comb-assisted Rydberg electrometry has enabled parallel access to multiple RF channels over an ultra-wide carrier-frequency range~\cite{MC-5}, while a multi-carrier RAQR employing a MW-frequency-comb local field has been proposed to extend the effective reception bandwidth for multicarrier waveforms, including orthogonal frequency-division multiplexing (OFDM)~\cite{huizhi_wang}. These efforts establish the feasibility of multicarrier reception with Rydberg atoms. However, existing approaches primarily employ the atomic medium as a broadband or multichannel front end, rather than proactively shaping its response for interference suppression. Tailoring the atomic response to the spectral structure of OFDM for direct inter-carrier interference (ICI) mitigation at the sensing stage remains largely unexplored.

OFDM imposes requirements that differ fundamentally from those of generic multicarrier reception. Its subcarrier orthogonality is highly sensitive to carrier-frequency offset (CFO) and oscillator phase noise (PN), both of which cause spectral leakage across neighboring subcarriers and consequently generate inter-carrier interference~\cite{Moose1994CFO,Tomba1998PN}. Conventional OFDM receivers therefore rely primarily on synchronization, CFO/PN estimation, and subsequent digital-domain compensation or ICI cancellation. Importantly, after fast Fourier transform (FFT) demodulation, the resulting interference exhibits a structured distribution over the discrete difference-frequency grid, with most of leakage typically concentrated around a limited number of neighboring subcarriers. This structure opens a distinct opportunity for a programmable Rydberg atomic front end. By shaping the relative responses between interfering-source and destination subcarriers according to the structured CFO/PN leakage weights, dominant inter-carrier leakage can be attenuated before digital demodulation. 
Such spectral shaping is not without cost, as increased selectivity may prolong the effective impulse response and intensify inter-symbol interference (ISI) induced by finite cyclic prefix (CP). This coupling between frequency-domain interference suppression and time-domain channel memory motivates a joint atomic-front-end design for OFDM reception.

Motivated by these challenges, our contributions are summarized as follows.
\begin{itemize}
\item
Building upon our previous work that the detected response of a single-LO SWM-based RAQR can be accurately approximated by an effective two-pole low-pass kernel~\cite{chen-Rydberg3}, we propose an atomic-comb Rydberg receiver tailored to OFDM communications. In particular, injecting a multi-tone LO generates frequency-shifted replicas of the intrinsic single-LO SWM response, with their spectral locations determined by the respective LO frequencies. These replicas coexist in the same atomic medium and combine coherently at the optical readout. By subtly adjusting the amplitude and phase of each LO tone, constructive and destructive interference can be selectively induced to engineer the atomic response components across the
source-destination subcarrier pairs involved in ICI coupling. 
This produces a configurable comb-like response, enabling the atomic front end to selectively shape OFDM-relevant frequencies without changing the downstream OFDM processing chain.

\item
The response-shaping capability of the proposed atomic comb is harnessed to proactively mitigate OFDM interference. Specifically, we demonstrate that CFO/PN-induced ICI follows a discrete difference-frequency structure, enabling the atomic-comb response to be shaped directly at the offsets responsible for inter-subcarrier leakage. Using the corresponding CFO/PN-dependent interference weights, the complex LO weights are jointly optimized to suppress dominant ICI components while preserving a prescribed reference response. We further establish the connection between the frequency-domain atomic-comb response and the resulting time-domain effective impulse response, revealing that ICI-oriented spectral shaping can modify the receiver memory and hence influence finite-CP-induced ISI. This motivates a joint ICI-ISI shaping framework that simultaneously controls both inter-carrier leakage and energy beyond the CP boundary, facilitating channel-memory shortening and reduced CP length.

\item
We numerically evaluate the SWM-based RAQR using the QuTiP toolkit~\cite{QuTip} and validate the proposed atomic-comb architecture through OFDM-level simulations. Our results reveal that, compared with the single-LO SWM-based RAQR, the proposed atomic comb can effectively reshape the receiver response over the OFDM difference-frequency grid, achieving an approximately 29.27\% reduction in average post-equalization ICI power over the considered CFO range. To evaluate the proposed joint ICI-ISI shaping strategy, we benchmark it against conventional RF reception and a 13-tap time-domain equalizer (TEQ), with the latter representing a conventional channel-shortening approach and using the same number of complex variables as the 13-tone atomic comb. Subject to the same residual-ISI constraint, the joint design reduces the required CP length from approximately 18 samples for both the conventional RF and TEQ baselines to 11 samples. This confirms that the atomic-comb shaping can simultaneously mitigate ICI, shorten the effective channel memory, and substantially lower CP overhead.
\end{itemize}


\section{Preliminary}
\label{sec2:preliminary}

This section briefly reviews the essential OFDM signal model and clarifies how a single-LO SWM-based RAQR can be abstracted as a linear time-invariant (LTI) two-pole low-pass front-end. 
These preliminaries establish a unified notation for the subsequent atomic-comb analysis.

\subsection{OFDM Baseband Signal Model and Cyclic Prefix}
\label{subsec:ofdm_model}

We consider an OFDM system with $N$ subcarriers and subcarrier spacing $\Delta f = 1/T_u$, where $T_u$ denotes the useful symbol duration. Let $\left\lbrace X_l\right\rbrace _{l=0}^{N-1}$ be the complex data symbols in one OFDM block. The continuous-time complex baseband waveform over $t\in\left[ 0,T_u\right) $ is
$x\left( t \right) = \sum\nolimits_{l = 0}^{N - 1} {{X_l}{e^{\jmath 2\pi l\Delta ft}}} $. To mitigate inter-symbol interference (ISI) due to channel delay spread, a CP of duration $T_{\mathrm{cp}}$ is appended, yielding the total OFDM block length $T = T_u + T_{\mathrm{cp}}$. At the receiver, the CP is removed and an $N$-point FFT is applied over the interval of length $T_u$.
Let $h \left( t \right) $ denote the equivalent baseband impulse response (including propagation channel and any front-end filtering), and let $ w \left(t \right) $ denote additive noise. By assuming perfect timing such that the FFT window is aligned after cyclic prefix (CP) removal, the received baseband signal can be given by $y\left( t \right) = \left( {h*x} \right)\left( t \right) + w\left( t \right),t \in \left[ {0,{T_u}} \right)$.
In the ideal synchronized case (no carrier offset and no phase noise), the FFT output on the $k$-th subcarrier becomes $Y_k = H_k X_k + W_k$, where $H_k$ denotes the discrete frequency response sampled at $f=k\Delta f$ and $W_k$ represents the noise term.

\begin{figure*}[t]
	\centering
	\includegraphics[width=0.9\textwidth]{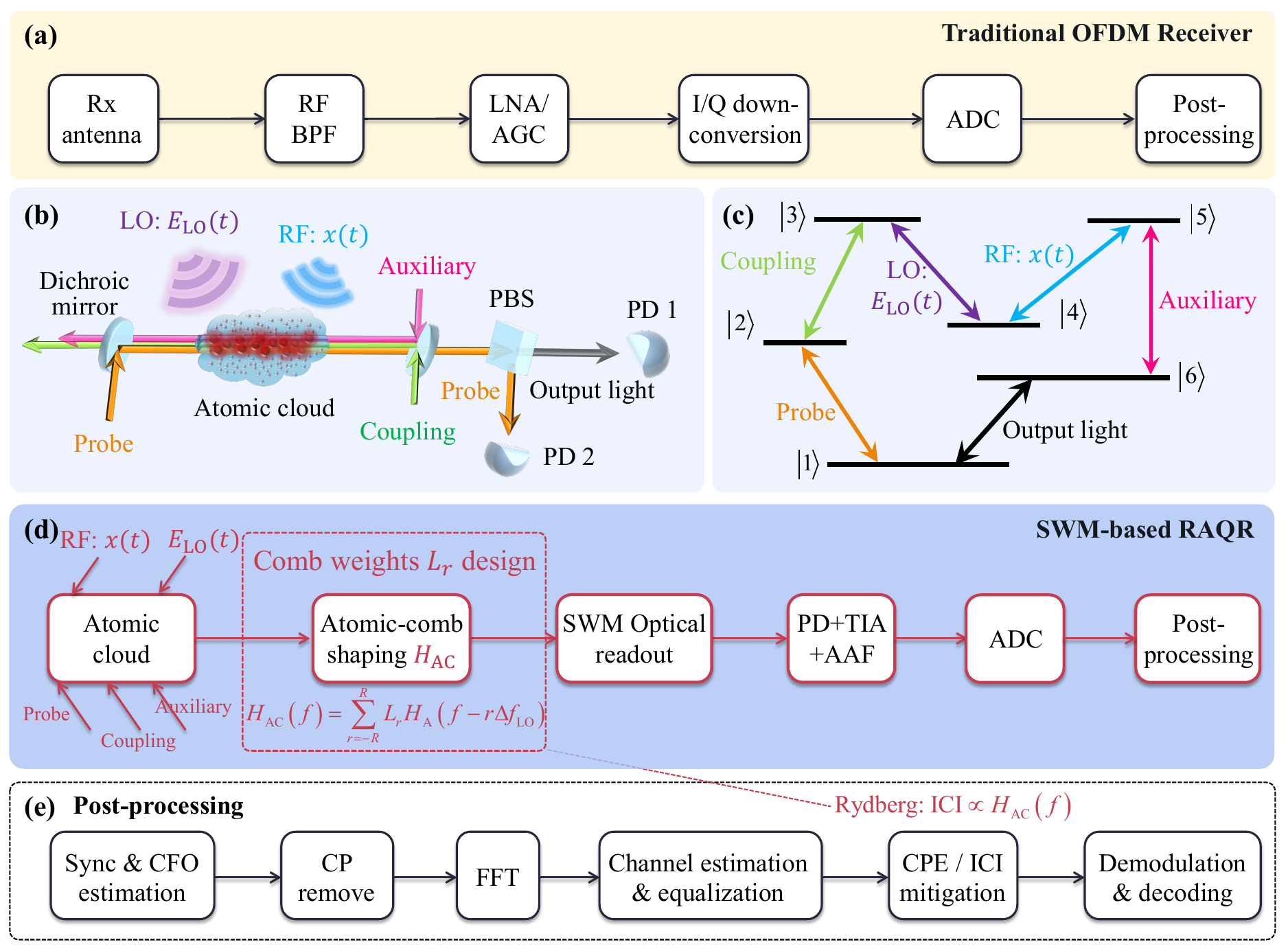}
	\caption{System-level block diagram. (a) Conventional electronic OFDM receiver. (b) Illustration of the SWM-based RAQR. (c) Six-level SWM transition configuration. (d) Proposed atomic-comb-enabled receiver based on SWM configuration. (e) Common OFDM baseband processing chain after digitization.		
		The abbreviations used in this figure are defined as: receiver (Rx), radio frequency (RF), band-pass filter (BPF), low-noise amplifier (LNA), automatic gain control (AGC), in-phase/quadrature (I/Q), analog-to-digital converter (ADC), photodetector (PD), transimpedance amplifier (TIA), anti-aliasing filter (AAF), synchronization (sync), carrier frequency offset (CFO), cyclic prefix (CP), fast Fourier transform (FFT), common phase error (CPE), and inter-carrier interference (ICI).} \label{Fig_block_diagram}
\end{figure*}

In practical receivers, the CFO and PN introduce a multiplicative phase term. Let $\Delta f_0$ denote the residual CFO and $\phi(t)$ the stochastic PN process. By defining $\theta \left( t \right) \triangleq 2\pi \Delta {f_0}t + \phi \left( t \right)$, a standard compact model for the post-CP received signal is given by 
\begin{equation}
y\left( t \right) = {e^{\jmath \theta \left( t \right)}}\left( {h*x} \right)\left( t \right) + w\left( t \right),t \in \left[ {0,{T_u}} \right).
\end{equation}
After FFT, the $k$-th subcarrier output can be expressed as
\begin{equation}\label{eq:fft_ici_general}
{Y_k} = \sum\limits_{l = 0}^{N - 1} {{H_l}{X_l}{C_{l - k}}}  + {W_k},k = 0,...,N - 1,
\end{equation}
where $C_n$ denotes the Fourier-series coefficients of $e^{\jmath\theta\left(t \right) }$ over the FFT window:
\begin{equation}\label{eq:Cn_def}
{C_n} \triangleq \frac{1}{{{T_u}}}\int_0^{{T_u}} {{e^{\jmath \theta \left( t \right)}}{e^{ - \jmath 2\pi n\Delta ft}}} {\mkern 1mu} dt,
\end{equation}
in which the term with $n=0$ corresponds to the common phase error (CPE), while the terms with $n\neq 0$ produce inter-carrier interference (ICI). Eq.~\eqref{eq:fft_ici_general} highlights a key structural fact the interference leakage from subcarrier $l$ to subcarrier $k$ is controlled by the coefficient indexed by $n=l-k$, i.e., by the difference frequency $\left( l-k \right) \Delta f$. Therefore, although CFO/PN are continuous-time impairments, their impact on OFDM demodulation manifests through the discrete set of coefficients $\left\lbrace C_n\right\rbrace $ associated with the discrete difference-frequency grid $\left\lbrace n\Delta f\right\rbrace $. This difference-frequency-driven structure motivates designing and analyzing the atomic-comb front-end response in terms of its samples on the OFDM grid and, more importantly, its behavior on the difference-frequency offsets that dominate ICI coupling.

\subsection{SWM-Based RAQR as an LTI Front-End}
\label{subsec:swm_two_pole}

We next outline how an SWM-based Rydberg atomic receiver can be incorporated into the OFDM model. In an SWM-based Rydberg atomic receiver, the incident RF field modulates atomic coherences and is transduced to an optical readout through nonlinear wave mixing. 
As a result, the atomic receiver can be abstracted as an effective LTI front-end characterized by a baseband transfer function $H_{\mathrm{A}} \left(f \right)  $~\cite{chen-Rydberg3}
\begin{equation}\label{eq:atomic_lti}
	Y_{\mathrm{A}}\left(f \right)  = H_{\mathrm{A}}\left(f \right)  X_{\mathrm{in}}\left(f \right)  + N_{\mathrm{A}}\left(f \right) ,
\end{equation}
where $X_{\mathrm{in}}\left(f \right)$ is the spectrum of the RF-to-baseband equivalent input, $Y_{\mathrm{A}}\left(f \right)$ characterizes the spectrum of the electrical readout, and $N_{\mathrm{A}}\left(f \right)$ captures the aggregated noise contributions~\cite{chen-Rydberg1}.
For the single-LO excitation configuration, the effective response of the SWM atomic receiver behaves as a two-pole low-pass system~\cite[Sec.~III-A]{chen-Rydberg3}
\begin{equation}\label{eq:two_pole_lp}
	H_{\mathrm{A}}(f) \approx \frac{G_{\mathrm{eq}}}{\left(1 + \jmath \frac{f}{f_{p1}}\right)\left(1 + \jmath \frac{f}{f_{p2}}\right)},
\end{equation}
where $G_{\mathrm{eq}}$ denotes the equivalent system response gain and $f_{p1},f_{p2}>0$ denote two characteristic pole frequencies determined by the underlying atomic dephasing rates and SWM dynamics. 
We then formulate a standard OFDM receiver model in which the overall frequency response on the $k$-th subcarrier becomes
\begin{equation}\label{eq:overall_Hl}
	H_k = H_{\mathrm{ch}}\left( k\Delta f\right)  H_{\mathrm{A}}\left( k\Delta f\right) ,
\end{equation}
with $H_{\mathrm{ch}}\left( \cdot\right) $ denoting the propagation channel response. Equation~\eqref{eq:overall_Hl} indicates that the SWM atomic receiver not only extends the usable bandwidth through tunable $H_{\mathrm{A}}\left(f \right)$, but also shapes the subcarrier-domain response that interacts with CFO/PN-induced coefficients $\left\lbrace C_n \right\rbrace $ in \eqref{eq:fft_ici_general}. This observation motivates the atomic-comb concept developed in the next section, where multi-tone LO excitation generalizes the current two-pole kernel into a comb-shaped front-end response to further control ICI in a difference-frequency-aware manner.

Fig.~\ref{Fig_block_diagram} illustrates the architectural contrast between a traditional OFDM receiver and the proposed SWM-based RAQR with atomic-comb shaping. 
As shown in Fig.~\ref{Fig_block_diagram}(a) for traditional OFDM receivers, the received OFDM RF waveform is processed by a classical analog front-end (RF BPF, LNA/AGC, I/Q down-conversion, ADC) and then by standard digital baseband modules (sync/CFO estimation, CP removal, FFT, channel estimation/equalization, and CPE/ICI mitigation, as summarized in Fig.~\ref{Fig_block_diagram}(e)). 
In our proposed architecture, presented in Fig.~\ref{Fig_block_diagram}(d), the information-bearing OFDM signal $x\left( t \right) $ (also known as the RF field) is coupled into a Rydberg vapor cell, while a locally generated multi-tone RF LO ${E_{{\text{LO}}}}\left( t \right)$ while optical probe, coupling and auxiliary fields drive the SWM transduction. This yields an atomic-comb response ${H_{{\text{AC}}}}\left( f \right) = \sum\nolimits_{r =  - R}^R {{L_r}} {H_{\text{A}}}\left( {f - r\Delta {f_{{\text{LO}}}}} \right)$, where the comb weights ${L_r}$ shape the front-end frequency selectivity. The SWM optical readout is subsequently detected by a PD and transimpedance amplifier (TIA) (with anti-aliasing filter (AAF)) and digitized by an ADC, after which the same standard OFDM baseband processing applies.
Importantly, by shaping ${H_{{\text{AC}}}}\left( f \right)$ to introduce attenuation at dominant discrete difference-frequency offsets, the atomic front-end can suppress ICI induced by CFO/phase-noise in a difference-frequency-aware manner. This reduces the burden of digital ICI mitigation while preserving compatibility with standard OFDM synchronization, equalization, and decoding stages.

\begin{figure*}[t]
	\centering
	\includegraphics[width=0.9\textwidth]{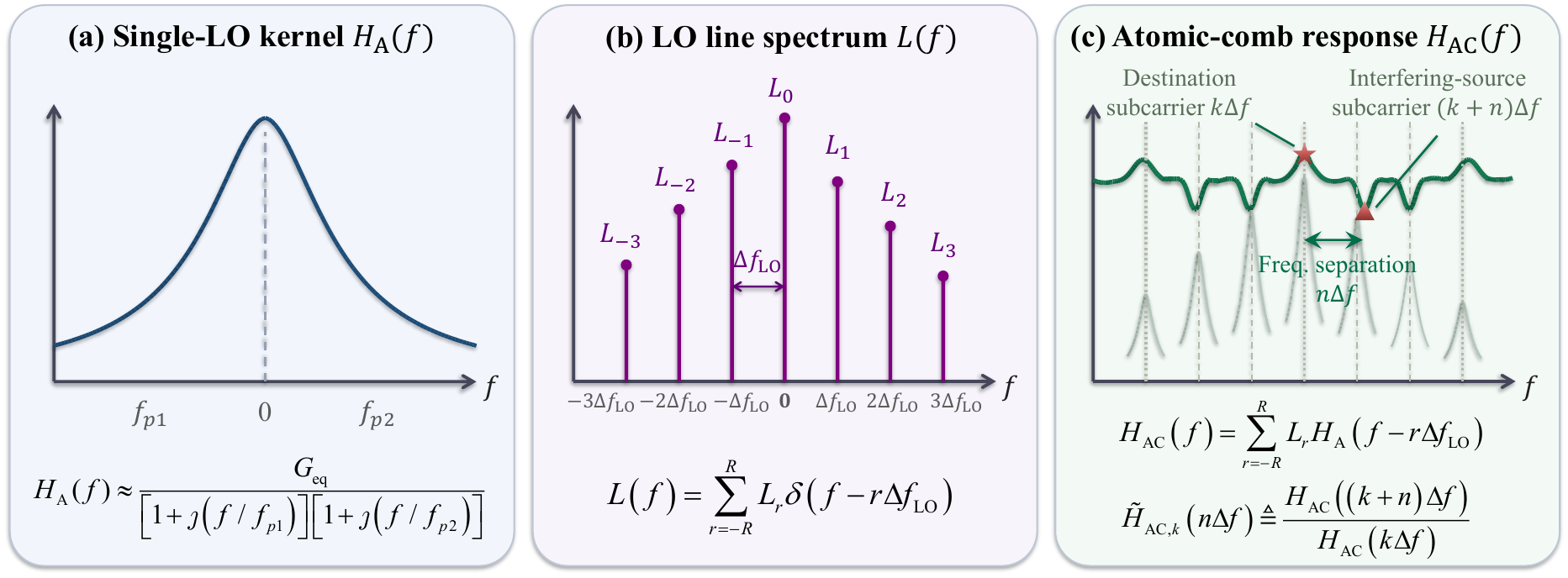}
	\caption{Illustration of the proposed atomic comb. (a) Single-LO kernel response $H_{\mathrm{A}}\left( f\right) $ of the SWM transduction with a pair of poles $f_{p1}$ and $f_{p2}$. (b) Discrete multi-tone injected LO line spectrum $L(f)$ with complex weights $\{L_r\}$. (c) Atomic-comb response $H_{\mathrm{AC}}\left( f\right) $ and a representative $n$-th-order ICI pair from $\left( k+n\right) \Delta f$ to $k\Delta f$, separated by $n\Delta f$. After one-tap equalization, the atomic-front-end coupling is characterized by $\widetilde H_{\mathrm{AC},k}\left( n\Delta f\right)$.
	} \label{atomic_comb}
\end{figure*}

\section{Atomic Comb}
\label{sec3:atomic_comb}


Building upon the single-LO configuration in Sec.~\ref{sec2:preliminary}, we now consider a multi-tone LO excitation to generalize the SWM-based RAQR into an atomic-comb front-end for the OFDM system. 



Let the LO field contain $2R+1$ tones with tone spacing $\Delta f_{\mathrm{LO}}$. Using the complex-envelope representation, the multi-tone LO can be written as
\begin{equation}\label{eq:LO_multitone}
	{E_{\mathrm{LO}}}\left( t \right) = \sum\limits_{r =  - R}^R {L_r} {e^{\jmath 2\pi r\Delta {f_{\mathrm{LO}}}t}},
\end{equation}
where $L_r\in\mathbb{C}$ denotes the complex weight (amplitude and phase) of the $r$-th LO tone. 
The SWM-based RAQR receives an information-bearing OFDM signal, while the multi-tone LO is a locally generated and injected reference used to control the atomic transduction response. The incident RF field is unknown and can be expressed by
\begin{equation}
 x\left( t \right) = \sum\limits_{l = 0}^{N - 1} {{X_l}{e^{\jmath 2\pi l\Delta ft}}} , 
\end{equation}
where $X_l$ carries the data symbols to be demodulated. By contrast, the RAQR applies a known multi-tone LO excitation to the vapor cell, whose complex weights $\left\lbrace L_r\right\rbrace $ are design variables that shape the effective atomic-comb response $H_{\mathrm{AC}}\left( f\right) $ during SWM transduction.
Under the small-signal regime, the overall RF-to-optical readout remains linear with respect to ${E_{{\text{RF}}}}\left( t \right)$, while the frequency response of this linear operator is parametrized by the injected LO tones $\left\lbrace L_r\right\rbrace $. Therefore, the atomic comb can be understood as an LO-programmable front-end shaping mechanism that functions on the received OFDM signal.

Recall that the RAQR employing a single LO admits an LTI transfer function $H_{\mathrm{A}}\left( f \right)$, as illustrated in Fig.~\ref{atomic_comb}(a), whose frequency selectivity is determined by the intrinsic SWM dynamics. When the injected LO contains multiple tones, each tone effectively generates a frequency-shifted copy of the same kernel response $H_{\mathrm{A}}\left( f \right)$. Therefore, the resultant atomic-comb front-end frequency response can be modeled as
\begin{equation}\label{eq:H_AC_def}
	{H_{\mathrm{AC}}}\left( f \right) = \sum\limits_{r =  - R}^R {L_r} {H_{\mathrm{A}}}\left( {f - r\Delta {f_{\mathrm{LO}}}} \right),
\end{equation}
where ${H_{\mathrm{AC}}}\left( f \right)$ denotes the atomic-comb response.

As depicted in Fig.~\ref{atomic_comb}(b), the injected multi-tone LO can be characterized by a discrete line spectrum
\begin{equation}
	L\left( f \right)\triangleq \sum_{r=-R}^{R} L_r \delta \left(f-r\Delta f_{\mathrm{LO}}\right),
\end{equation}
where the complex coefficients $\left\lbrace L_r\right\rbrace $ are controllable design variables. The core idea is that the overall front-end response is amalgamated by the weighted superposition of these frequency-shifted kernel copies, yielding the comb-shaped response in Fig.~\ref{atomic_comb}(c), consistent with \eqref{eq:H_AC_def}. More explicitly, Fig.~\ref{atomic_comb}(c) illustrates
a representative source-destination subcarrier pair associated with
ICI leakage from an interfering-source subcarrier at $\left( k+n\right) \Delta f$ to a destination subcarrier at $k\Delta f$. Here, $n$ denotes the
difference-frequency order, while $n\Delta f$ represents the frequency
separation between the two subcarriers. For further insight, \eqref{eq:H_AC_def} admits an equivalent convolution interpretation
\begin{equation}
	H_{\mathrm{AC}}\left( f \right)=\left(L * H_{\mathrm{A}}\right)\left( f \right),
\end{equation}
which clarifies that the atomic-comb response is generated by shifting the same kernel $H_{\mathrm{A}}\left( f\right) $ to the LO-tone locations $r\Delta f_{\mathrm{LO}}$ and coherently weighting these shifted replicas by the complex coefficients $\{L_r\}$.

More importantly, CFO/PN-induced ICI is indexed by the difference-frequency order $n$, whereas the atomic front-end contribution depends on the relative response between the interfering-source subcarrier $k+n$ and the
destination subcarrier $k$. Therefore, the relevant front-end factor is
$H_{\mathrm{AC}}\left(  \left( k+n\right) \Delta f \right) / H_{\mathrm{AC}}\left( k\Delta f\right) $, rather than an isolated response sample at $n\Delta f$. This observation motivates the exact all-subcarrier ICI characterization developed in the next section.


\section{ICI Suppression via Atomic Comb}
\label{sec4:ici_suppression}


This section characterizes how the atomic-comb response affects CFO/PN-induced OFDM ICI and formulates the corresponding multi-tone LO design. The difference-frequency index determines the impairment-induced leakage weight, whereas the front-end contribution is governed by the relative response across the source and destination
subcarriers.


\subsection{Exact Post-Equalization ICI Characterization}
\label{subsec:ici_mechanism}

Recall that the FFT output on the $k$-th subcarrier under residual CFO/PN can be expressed by \eqref{eq:fft_ici_general}. Separating the desired term and the ICI term yields
\begin{equation}\label{eq:Yk_decomp}
{Y_k} = {H_k}{X_k}{C_0} + \underbrace {\sum\nolimits_{l \ne k} {{H_l}{X_l}{\mkern 1mu} {C_{l - k}}} }_{ \triangleq {\kern 1pt} {I_k}} + {W_k},
\end{equation}
where $I_k$ denotes the ICI contribution and $H_k = H_{\mathrm{ch}}(k\Delta f) H_{\mathrm{AC}}\left( k\Delta f\right) $ represents the overall response on the $k$-th subcarrier.

\subsubsection{Difference-Frequency Structure}
\label{subsubsec:diff_structure}
By re-indexing using the difference-frequency index $n= l-k$, the ICI term can be equivalently written as
\begin{equation}\label{eq:Ik_diff}
	I_k=\sum_{n\neq 0} H_{k+n}  X_{k+n}  C_{n},
\end{equation}
which explicitly indicates that each ICI component is indexed by the
difference-frequency order $n$, corresponding to a frequency separation
of $n\Delta f$ between the source and destination subcarriers.

\subsubsection{Post-Equalization ICI and the Role of Atomic Comb}
\label{subsubsec:post_eq_ici}

To make the ICI-suppression mechanism transparent, we consider a standard one-tap per-subcarrier equalization that compensates the desired-path gain. Specifically, we define
\begin{equation}\label{eq:one_tap_eq}
	\widehat{X}_k \triangleq \frac{Y_k}{H_k C_0}.
\end{equation}
Substituting \eqref{eq:Yk_decomp} into \eqref{eq:one_tap_eq} yields
\begin{equation}\label{eq:Xhat_decomp}
{{\widehat X}_k} = {X_k} + \underbrace {\sum\nolimits_{n \ne 0} {\frac{{{H_{k + n}}}}{{{H_k}}}\frac{{{C_n}}}{{{C_0}}}{X_{k + n}}} }_{ \triangleq {{\widehat I}_k}} + {{\widehat W}_k},
\end{equation}
where $\hat{I}_k$ denotes the post-equalization ICI term and $\widehat{W}_k \triangleq W_k/\left( H_k C_0\right) $ is the equalized noise.
Equation~\eqref{eq:Xhat_decomp} indicates that the effective ICI coupling coefficient for the $n$-th difference-frequency order is the product of two factors: (i) the CFO/PN factor $C_n/C_0$ and (ii) the relative end-to-end frequency response ratio $H_{k+n}/H_k$.
Then, the ratio $H_{k+n}/H_k$ can be factorized as
\begin{equation}\label{eq:Hratio_decomp}
\frac{{{H_{k + n}}}}{{{H_k}}} = \frac{{{H_{{\text{ch}}}}\left( {\left( {k + n} \right)\Delta f} \right)}}{{{H_{{\text{ch}}}}\left( {k\Delta f} \right)}} \times \frac{{{H_{{\text{AC}}}}\left( {\left( {k + n} \right)\Delta f} \right)}}{{{H_{{\text{AC}}}}\left( {k\Delta f} \right)}}.
\end{equation}

For a given data subcarrier $k\in\mathcal{K}_{\mathrm d}$, we define $\mathcal{N}_{k}\triangleq
\{ n \in\mathbb{Z}\setminus\{0\}:k+n\in\mathcal{K}_{\mathrm d}\}$ as the set of admissible ICI orders. Unlike the CFO/PN coefficient $C_n/C_0$, which is indexed only by the difference-frequency order $n$, the atomic-comb contribution in \eqref{eq:Hratio_decomp} generally depends on both $k$ and $n$. We therefore define the $k$-dependent normalized atomic-comb ratio as
\begin{equation}\label{tilde_H_AC}
	\widetilde{H}_{\mathrm{AC},k}\left( n\Delta f\right) 
	\triangleq
	\frac{
		H_{\mathrm{AC}}\left(  \left( k+n\right) \Delta f \right) 
	}{
		H_{\mathrm{AC}}\left( k\Delta f\right) 
	},
	 n\in\mathcal{N}_{k}
\end{equation}
which characterizes how the atomic front end modifies the leakage from subcarrier $k+n$ to subcarrier $k$ after one-tap equalization. In general, it cannot be reduced to a function of
$n$ alone.

We assume $\left\lbrace X_k \right\rbrace$ are i.i.d. with $\mathbb{E}\left[ {{{\left| {{X_k}} \right|}^2}} \right] = {P_X}$ and are independent of $\left\lbrace C_n \right\rbrace$. Define the normalized CFO/PN coefficients ${\bar{C}}_n\triangleq C_n / C_0$ and their second-order statistics
\begin{equation}\label{eq:wn_def}
	{w_n} \triangleq \mathbb{E}\left[ {{{\left| {\bar{C}_n} \right|}^2}} \right].
\end{equation}
Conditioned on a given channel realization, the normalized post-equalization ICI power on subcarrier $k$ is then given by
\begin{align}
{\mathcal{J}_{{\text{ICI}},k}} & \triangleq \mathbb{E}\left[ {{{\left| {{{\hat I}_k}} \right|}^2}} \right]/{P_X}
\nonumber\\
&= \sum\limits_{n \in {\mathcal{N}_k}} {{w_n}{{\left| {\frac{{{H_{{\text{ch}}}}\left( {\left( {k + n} \right)\Delta f} \right)}}{{{H_{{\text{ch}}}}\left( {k\Delta f} \right)}}} \right|}^2}{{\left| {{{\tilde H}_{{\text{AC}},k}}\left( {n\Delta f} \right)} \right|}^2}} .
\end{align}
Accordingly, the exact average post-equalization ICI metric over the active data subcarriers is formulated by
\begin{equation}\label{J_ICI}
{\mathcal{J}_{{\text{ICI}}}} \triangleq \frac{1}{{\left| {{\mathcal{K}_{\text{d}}}} \right|}}\sum\limits_{k \in {\mathcal{K}_{\text{d}}}} {{\mathcal{J}_{{\text{ICI}},k}}} .
\end{equation}
For a flat propagation channel, the channel-response ratio reduces to $H_\text{ch} \left( \cdot \right) = 1 $. Nevertheless, the
atomic-comb ratio remains dependent on $k$ unless additional
structure is imposed on $H_{\mathrm{AC}}\left( f\right) $.

Eqs.~\eqref{tilde_H_AC}-\eqref{J_ICI} also clarify the roles of the two frequency indices. The index $n$ identifies the CFO/PN-induced difference-frequency order through $w_n$, whereas the front-end contribution is determined by the response ratio between the actual subcarrier frequencies $\left( k+n\right) \Delta f$ and $k\Delta f$. Hence, the points $\{n\Delta f\}$ do not form a separate physical grid from the active-subcarrier samples.
In particular,
$H_{\mathrm{AC}}\left( n\Delta f\right) /H_{\mathrm{AC}}\left( 0\right) $ represents
only the special case $k=0$ and cannot characterize the ICI experienced by all data subcarriers. Consequently, imposing deep attenuation at $H_{\mathrm{AC}}\left( n\Delta f\right) $ may reduce the interference
entering the reference subcarrier $k=0$, but it may
simultaneously attenuate the desired signal carried by the $n$-th subcarrier. Moreover, if the atomic-comb magnitude is strictly flat over all active subcarriers, then $\left| \widetilde{H}_{\mathrm{AC},k}\left( n\Delta f\right) \right| \approx 1$, leaving little scope for suppressing active-to-active ICI through magnitude shaping alone. The atomic-comb design should therefore balance the exact all-subcarrier ICI metric in \eqref{J_ICI} against controlled desired-gain variation, rather than simultaneously claiming a globally flat passband and independent attenuation at the same active-subcarrier frequencies.

Since the denominators $H_{\mathrm{AC}}\left( k\Delta f\right) $ also depend on the multi-tone LO weights, directly minimizing \eqref{J_ICI} generally leads to a nonconvex fractional optimization problem. The next subsection develops a tractable atomic-comb design while explicitly distinguishing any low-complexity surrogate from the exact all-subcarrier metric in \eqref{J_ICI}.

\subsection{Atomic-Comb Shaping}
\label{subsec:atomic_comb_shaping}

We next exploit the multi-tone LO weights $\{L_r\}$ to control the exact all-subcarrier ICI metric derived in~\eqref{J_ICI}. Different from the previous pointwise formulation, the design does not directly suppress the absolute samples
$H_{\mathrm{AC}} \left( n\Delta f\right) $. Instead, it optimizes the relative responses between every interfering-source subcarrier $k+n$ and its destination subcarrier $k$.

By stacking the multi-tone LO weights into a vector, we define
$	{\mathbf{L}} = {\left[ {{L_{ - R}},{L_{ - R + 1}},...,{L_0},...,{L_R}} \right]^T} \in {\mathbb{C}^{\left( {2R + 1} \right) \times 1}}$.
For any frequency $f$, we define the kernel vector
\begin{align}\label{eq:bvec_def}
{\mathbf{b}}\left( f \right) \triangleq & \left[ {{H_{\text{A}}}\left( {f + R\Delta {f_{{\text{LO}}}}} \right),{H_{\text{A}}}\left( {f + (R - 1)\Delta {f_{{\text{LO}}}}} \right),...,} \right. \nonumber\\
 & \ {\left. {{H_{\text{A}}}\left( {f - r\Delta {f_{{\text{LO}}}}} \right),...,{\mkern 1mu} {H_{\text{A}}}\left( {f - R\Delta {f_{{\text{LO}}}}} \right)} \right]^H}
\end{align}
whose entries are ordered according to the LO tone index $r=-R,-R+1,...,R$, and the entry associated with tone index $r$ corresponds to the shifted kernel $H_{\mathrm{A}}\left( f-r\Delta f_{\mathrm{LO}}\right) $.
Then \eqref{eq:H_AC_def} can be written compactly as
\begin{equation}\label{eq:HAC_vec}
	{H_{{\text{AC}}}}\left( f \right) = {{\mathbf{b}}^H}\left( f \right){\mathbf{L}}.
\end{equation}
Upon letting $\mathbf{b}_{q}\triangleq\mathbf{b}\left( q\Delta f\right) $, the $k$-dependent response ratio in~\eqref{tilde_H_AC} is given by
\begin{equation}
	\widetilde{H}_{\mathrm{AC},k}\left( n\Delta f\right) 
	=
	\frac{\mathbf{b}_{k+n}^{H}\mathbf{L}}
	{\mathbf{b}_{k}^{H}\mathbf{L}}, n\in\mathcal{N}_{k}.
\end{equation}
Given $\mathcal{J}_{\mathrm{ICI}}$ in~\eqref{J_ICI}, the
atomic-comb design can be formulated as
\begin{subequations}
	\begin{align}
		\underset{\mathbf{L}}{\operatorname{minimize}}
		\quad&
		\mathcal{J}_{\mathrm{ICI}}
		\label{eq:exact_ici_design_obj}
		\\
		\operatorname{s.t.}\quad&
		\mathbf{b}_{0}^{H}\mathbf{L}=1,
		\label{eq:exact_ici_design_ref}
		\\
		&
		\|\mathbf{L}\|_{2}^{2}\leq P_{\max}.
		\label{eq:exact_ici_design_power}
	\end{align}\label{optimization_ICI}
\end{subequations}

\noindent In~\eqref{optimization_ICI}, the constraint \eqref{eq:exact_ici_design_ref} fixes the otherwise arbitrary overall scale and phase of the atomic-comb response, while~\eqref{eq:exact_ici_design_power} constrains the total multi-tone LO weight. Since $\mathcal{J}_{\mathrm{ICI}}$ accounts for every
$k\in\mathcal{K}_{\mathrm d}$, attenuating a particular active subcarrier generally increases the corresponding post-equalization response ratios and is therefore penalized by the objective itself. 
Problem~\eqref{optimization_ICI} is a nonconvex fractional program and can be solved via successive convex approximation (SCA)~\cite{SCA}, where the denominator terms are locally lower-bounded by affine functions, resulting in a convex quadratic-over-linear subproblem at each iteration until convergence to a stationary point.

\begin{figure*}[t]
	\centering
	\includegraphics[width=\textwidth]{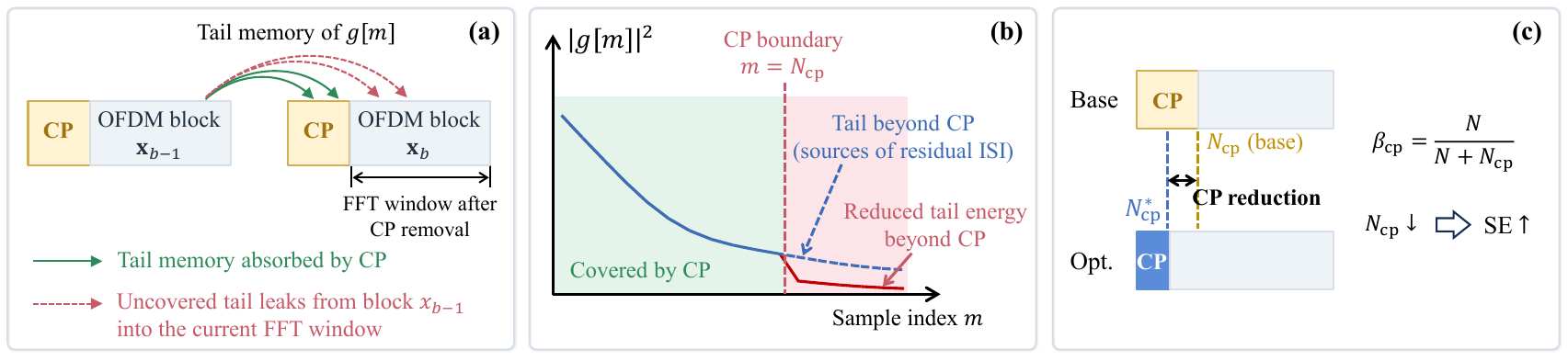}
	\caption{Illustration of atomic-comb-enabled ISI-aware shaping and CP reduction. (a) Insufficient CP induces leakage. (b) Atomic-comb shaping reduces the post-CP tail. (c) The optimized atomic comb shortens the effective memory sufficiently, thus reducing CP overhead and improving SE.
	} \label{three_panel}
\end{figure*}

\section{Extension to ISI-Aware Shaping and CP Reduction}\label{sec5:ISI-CP}

Sec.~\ref{sec4:ici_suppression} has established atomic-comb shaping from the perspective of CFO/PN-induced ICI suppression, while this section asks whether the same atomic-comb degrees of freedom can also shape the time-domain effective memory of the impulse response.

\subsection{Effective Impulse Response and CP Insufficiency}

We commence by revisiting the effective impulse response of the OFDM link under the proposed atomic-comb front-end. Recall that, with the propagation channel $H_{\mathrm{ch}}$ and the atomic-comb response $H_\mathrm{AC}$ combined, the overall subcarrier-domain response can be expressed as $H_k = H_{\mathrm{ch}}\left( k\Delta f\right)  H_{\mathrm{AC}}\left( k\Delta f\right) $. Accordingly, the corresponding discrete-time effective impulse response, denoted by $g\left[m\right]$,  can be defined through the inverse discrete Fourier transform (IDFT) as
\begin{equation}\label{eq:g_idft}
g\left[ m \right] \triangleq \frac{1}{N}\sum\limits_{k = 0}^{N - 1} {{H_k}{e^{\jmath \frac{{2\pi }}{N}km}}} ,m = 0,1,...,N - 1.
\end{equation}
For ease of derivations, let $L_g$ denote a sufficiently large truncation length such that the residual energy beyond $m=L_g-1$ is negligible. {Since $H_k$ already contains the atomic-comb response $H_\mathrm{AC} \left(k \Delta f \right) $, the effective impulse response $g \left[m\right]$ is not merely a channel-induced quantity, but an atomic-front-end-shaped memory profile.}

Next, we denote the $b$-th OFDM block by ${{\mathbf{x}}_b} = {\left[ {{x_b}\left[ 0 \right],{x_b}\left[ 1 \right],...,{x_b}\left[ {N - 1} \right]} \right]^T}$,
with CP length $N_{\mathrm{cp}}$. After CP insertion, transmission over the equivalent channel \eqref{eq:g_idft}, and CP removal at the receiver, the $n$-th time-domain sample within the FFT window can be expressed as
\begin{equation}\label{eq:y_postcp_basic}
{y_b}\left[ n \right] = \sum\limits_{m = 0}^{{L_g} - 1} {g\left[ m \right]{{\tilde x}_b}\left[ {n - m} \right] + {w_b}\left[ n \right]} ,n = 0,1, \ldots ,N - 1,
\end{equation}
where ${{{\tilde x}_b}\left[ \cdot \right]}$ denotes the CP-extended transmit sequence and ${{w_b}\left[ n \right]}$ is the effective noise sample. When the CP is sufficiently long relative to the effective memory of ${g\left[ m \right]}$, the linear convolution in \eqref{eq:y_postcp_basic} is converted into a circular convolution over the useful OFDM block. However, if the tail of ${g\left[ m \right]}$ extends beyond the CP duration, the received block is contaminated by samples from the previous OFDM symbol, and the circular-convolution property no longer holds exactly. As illustrated in Fig.~\ref{three_panel}(a), after CP insertion and transmission over $g\left[m\right]$, the portion of the memory that remains within the CP support is absorbed and still behaves circularly after CP removal, whereas the uncovered tail extends across OFDM blocks and leaks from $\mathbf{x}_{b-1}$ into the current FFT window associated with $\mathbf{x}_b$. 

Furthermore, \eqref{eq:y_postcp_basic} can be decomposed as
\begin{equation}\label{eq:y_cp_decomp}
{y_b}\left[ n \right] = \underbrace {\sum\limits_{m = 0}^{{L_g} - 1} {g\left[ m \right]{x_b}\left[ {{{\left\langle {n - m} \right\rangle }_N}} \right]} }_{{\text{Ideal circular term}}} + {e_b}\left[ n \right] + {w_b}\left[ n \right],
\end{equation}
where $\langle \cdot \rangle_N$ denotes modulo-$N$ operation and ${e_b}\left[ n \right]$ collects the residual distortion caused by insufficient CP coverage, which is given by
\begin{align}\label{eq:e_cp_insufficient}
		\resizebox{0.48\textwidth}{!}{$
		{e_b}\left[ n \right] = \sum\limits_{m = {N_{{\text{cp}}}} + n + 1}^{{L_g} - 1} {g\left[ m \right]\left( {{x_{b - 1}}\left[ {N + {N_{{\text{cp}}}} + n - m} \right] - {x_b}\left[ {{{\left\langle {n - m} \right\rangle }_N}} \right]} \right)} .$}	
\end{align}
Eq.~\eqref{eq:e_cp_insufficient} implies that the terms with $m \le N_{\mathrm{cp}}+n$ are still absorbed by the CP and hence behave circularly, whereas the terms with $m > N_{\mathrm{cp}}+n$ exceed the CP support and therefore introduce inter-block leakage. Consequently, the tail of ${g\left[ m \right]}$ outside the CP is the fundamental source of residual ISI/IBI and of the subsequent additional interference observed after FFT demodulation.
Additionally, Fig.~\ref{three_panel}(b) provides an energy-domain view.
The CP boundary at $m=N_{\mathrm{cp}}$ separates the portion of ${{{\left| {g\left[ m \right]} \right|}^2}}$ covered by the CP from the residual tail lying beyond it.
Only the latter contributes to post-CP interference, and therefore reducing the tail energy beyond the CP boundary is the key goal of ISI-aware shaping.

Based on~both~\eqref{eq:y_cp_decomp} and~\eqref{eq:e_cp_insufficient}, the residual ISI-to-signal ratio is defined as 
\begin{equation}\label{eta_ISI}
{\eta _{{\text{ISI}}}}\left( {{\mathbf{L}},{N_{{\text{cp}}}}} \right) \triangleq \frac{{\sum\nolimits_{n = 0}^{N - 1} \mathbb{E} \left[ {{{\left| {{e_b}\left[ n \right]} \right|}^2}} \right]}}{{\sum\limits_{n = 0}^{N - 1} {\mathbb{E}\left[ {{{\left| {\sum\limits_{m = 0}^{{L_g} - 1} {g\left[ m \right]} {x_b}\left[ {{{\left\langle {n - m} \right\rangle }_N}} \right]} \right|}^2}} \right]} }}
\end{equation}
The numerator measures the residual distortion caused by the portion of the effective impulse response not covered by the CP, whereas the denominator represents the useful within-block signal power. Compared with an unnormalized
tail-energy metric, \eqref{eta_ISI} prevents an apparent ISI reduction obtained merely by decreasing the overall effective-response gain.

When the time-domain OFDM samples are temporally uncorrelated and satisfy
$\mathbb{E}\left[ {\left| {{x_b}{{\left[ n \right]}^2}} \right|} \right] = {P_x}$, \eqref{eta_ISI} reduces to
\begin{equation}
{\eta _{{\text{ISI}}}}\left( {{\mathbf{L}},{N_{{\text{cp}}}}} \right) = \frac{{2\sum\nolimits_{m = {N_{{\text{cp}}}} + 1}^{{L_g} - 1} {{\alpha _m}} {{\left| {g\left[ m \right]} \right|}^2}}}{{\sum\nolimits_{m = 0}^{{L_g} - 1} {{{\left| {g\left[ m \right]} \right|}^2}} }},
	\label{eq:isi_ratio_white}
\end{equation}
where $\alpha_m\triangleq\frac{m-N_{\mathrm{cp}}}{N}$.
Hence, the proposed metric retains the weighted-tail interpretation of the original formulation, but normalizes the uncovered tail energy by the total useful response energy. 

\subsection{Joint ICI-ISI Atomic-Comb Shaping}

We next show that the tail of the effective impulse response can be explicitly expressed as a function of the multi-tone LO weight vector $\mathbf{L}$, thereby enabling a joint ICI-ISI atomic-comb design. We restructure the subcarrier-domain atomic-comb response $H_k$ as
\begin{align}
{H_k} &= {H_{{\text{ch}}}}\left( {k\Delta f} \right){H_{{\text{AC}}}}\left( {k\Delta f} \right) \nonumber\\
 &= {H_{{\text{ch}}}}\left( {k\Delta f} \right)\sum\limits_{r =  - R}^R {{L_r}{H_{{\text{A}}}}\left( {k\Delta f - r\Delta {f_{{\text{LO}}}}} \right)} \nonumber\\
  &= \sum\limits_{r =  - R}^R {{L_r}{\Psi _r}\left[ k \right]}, 
\end{align}
where ${\Psi _r}\left[ k \right] = {H_{{\text{ch}}}}\left( {k\Delta f} \right){H_{{\text{A}}}}\left( {k\Delta f - r\Delta {f_{{\text{LO}}}}} \right)$ connotes that the overall subcarrier response is affine in the LO weights $\left\lbrace L_r\right\rbrace $.
Applying the IDFT to ${\Psi _r}\left[ k \right]$, the effective impulse response $g\left[ m \right]$ can be decomposed as
\begin{equation}\label{eq:g_sum_gr}
g\left[ m \right] = \sum\limits_{r =  - R}^R {{L_r}{g_r}\left[ m \right]} ,m = 0,1, \ldots ,{L_g} - 1,
\end{equation}
where $	{g_r}\left[ m \right] \triangleq \frac{1}{N}\sum\nolimits_{k = 0}^{N - 1} {{\Psi _r}\left[ k \right]{e^{\jmath \frac{{2\pi }}{N}km}}} $ denotes the contribution of the $r$-th LO tone to the effective impulse response $g\left[ m \right]$. Hence, $g\left[ m \right]$'s tail is linearly parameterized by the vector of LO weights. Since both the useful circular-convolution component in~\eqref{eq:y_cp_decomp} and the residual distortion $e_b\left[ n \right] $ in~\eqref{eq:e_cp_insufficient} are linear in $g\left[ m\right] $, their average powers are quadratic in $\mathbf L$. Therefore, the residual ISI-to-signal ratio in~\eqref{eta_ISI} can be expressed as
\begin{equation}
{\eta _{{\text{ISI}}}}\left( {{\mathbf{L}},{N_{{\text{cp}}}}} \right) = \frac{{{{\mathbf{L}}^H}{{\mathbf{Q}}_{{\text{ISI}}}}\left( {{N_{{\text{cp}}}}} \right){\mathbf{L}}}}{{\frac{1}{{\left| {{\mathcal{K}_{\text{d}}}} \right|}}\sum\nolimits_{k \in {\mathcal{K}_{\text{d}}}} {{{\left| {{H_{{\text{ch}}}}\left( {k\Delta f} \right){\mathbf{b}}_k^H{\mathbf{L}}} \right|}^2}} }}.
	\label{eq:isi_quadratic_ratio}
\end{equation}
In~\eqref{eq:isi_quadratic_ratio}, $\mathbf{Q}_{\mathrm{ISI}}(N_{\mathrm{cp}})\succeq \mathbf{0}$ characterizes the residual distortion caused by insufficient CP coverage. The denominator represents the average useful received-signal power over the active subcarriers. This normalization prevents an apparent ISI reduction caused merely by decreasing the overall atomic-comb response.

We then formulate the joint ICI-ISI atomic-comb design as
\begin{subequations}\label{optimization-joint}
	\begin{align}
		\underset{\mathbf L}{\operatorname{minimize}}
		\quad&
		\eta_{\mathrm{ISI}}(\mathbf L,N_{\mathrm{cp}})
		\label{eq:joint_design_obj}
		\\
		\operatorname{s.t.}\quad&
		\mathcal{J}_{\mathrm{ICI}}\left( \mathbf L\right) 
		\leq
		\mathcal{J}_{\mathrm{ICI}}^{\mathrm{S\text{-}L}},
		\label{eq:joint_design_ici}
		\\
		&
		\mathbf b_0^{H}\mathbf L=1,
		\label{eq:joint_design_ref}
		\\
		&
		\|\mathbf L\|_2^2\leq P_{\max},
		\label{eq:joint_design_power}
	\end{align}
\end{subequations}
where $\mathcal{J}_{\mathrm{ICI}}\left( \mathbf L\right) $ is the exact average post-equalization ICI metric defined in~\eqref{J_ICI}, and $J_{\mathrm{ICI}}^{\mathrm{S\text{-}L}}$ denotes the corresponding ICI value of the single-LO SWM receiver. Therefore, \eqref{eq:joint_design_ici} ensures that the time-domain memory reduction is not achieved at the expense of degraded ICI performance. Unlike a weighted-sum formulation, the proposed design does not require an  arbitrary weighting factor between the ICI and ISI metrics.
Because both $J_{\mathrm{ICI}}\left( \mathbf L\right) $ and $\eta_{\mathrm{ISI}}\left( \mathbf L,N_{\mathrm{cp}}\right) $ contain
$\mathbf{L}$-dependent denominators, the resulting problem is nonconvex and can be solved using the same SCA procedure as problem~\eqref{optimization_ICI}.
Additionally, the formulation has a direct physical interpretation. The objective suppresses the portion of the effective channel memory that remains uncovered by the target CP, while the ICI constraint guarantees that the exact all-subcarrier ICI does not exceed that of the Single-LO reference. Thus, the same multi-tone LO weights jointly control frequency-domain inter-carrier leakage and time-domain post-CP interference.

\subsection{CP-Reduction and SE Implications}

A CP length $N_{\mathrm{cp}}$ is regarded as feasible if there exists an LO-weight vector $\mathbf L$ satisfying the constraints in~\eqref{eq:joint_design_ici}-\eqref{eq:joint_design_power} and
$	\eta_{\mathrm{ISI}}(\mathbf L,N_{\mathrm{cp}}) \leq \epsilon_{\mathrm{ISI}}$, where $\epsilon_{\mathrm{ISI}}$ denotes the prescribed residual ISI-to-signal threshold. Accordingly, the minimum feasible CP length is
\begin{equation}
	N_{\mathrm{cp}}^{\star}
	\triangleq
	\min \left\{
	N_{\mathrm{cp}}:
	\eta_{\mathrm{ISI}}
	\left( \mathbf L^{\star},N_{\mathrm{cp}}\right) 
	\leq\epsilon_{\mathrm{ISI}}
	\right\},
\end{equation}
where the optimal $\mathbf L^{\star}$ is obtained from problem~\eqref{optimization-joint}. Since~\eqref{eq:joint_design_ici} already constrains the exact ICI to be no greater than that of the Single-LO receiver, a smaller $N_{\mathrm{cp}}^{\star}$ directly indicates that the optimized atomic comb supports CP reduction without degrading the reference ICI performance.

For an OFDM block with useful length $N$ and CP length $N_{\mathrm{cp}}$, the CP-overhead factor, denoted by $\beta_{\mathrm{cp}}$, is defined as
$	\beta_{\mathrm{cp}}
	\triangleq
	\frac{N}{N+N_{\mathrm{cp}}}$.
It is intuitive that reducing the CP length directly increases the fraction of useful symbols per OFDM block, as illustrated in Fig.~\ref{three_panel}(c).
The atomic-comb design may also improve the effective received SINR by suppressing CFO/PN-induced ICI and post-CP tail interference. Consequently, the proposed joint shaping can improve the effective SE through two complementary mechanisms: reducing residual interference and lowering CP overhead.

\section{Simulation Results}\label{sec6:simulation}

\subsection{Simulation Configuration}
\label{subsec:sim_config}

This section specifies the simulation configuration adopted to evaluate the proposed SWM-based RAQR for OFDM signal reception. 
To retain the quantum-physical grounding of the proposed receiver model, the effective single-LO kernel used in this paper is instantiated from a concrete SWM-based Rydberg atomic operating point~\cite{chen-Rydberg3,SWM-2,SWM-2.5}. We adopt the same physically plausible six-level $^{87}\mathrm{Rb}$ SWM configuration as in our previous study~\cite{chen-Rydberg3}, with numerical calculations carried out by employing the QuTiP toolkit~\cite{QuTip}.
Specifically, we consider a cold atomic cloud with peak density on the order of $ 2\times10^{10}~\mathrm{cm}^{-3}$ and temperature around $70~\mu\mathrm{K}$. The probe, coupling, and auxiliary optical beams are focused onto the atomic ensemble, and the MW LO and RF fields are applied around $84.18~\mathrm{GHz}$ and $83.72~\mathrm{GHz}$, respectively.
The underlying atomic system is built on a six-level $^{87}\mathrm{Rb}$ manifold driven by five near-resonant fields, namely the probe (P), coupling (C), auxiliary (A), local oscillator (LO), and the incident RF signal. The corresponding nominal Rabi frequencies are chosen as
$\Omega_{\mathrm{P}}/2\pi = 1.14~\mathrm{MHz}$, $\Omega_{\mathrm{C}}/2\pi = 9.0~\mathrm{MHz}$, $\Omega_{\mathrm{A}}/2\pi = 6.2~\mathrm{MHz}$, $\Omega_{\mathrm{LO}}/2\pi = 1.4~\mathrm{MHz}$, and 
$\Omega_{\mathrm{RF}}/2\pi = 1.0~\mathrm{MHz}$.
The dominant coherence dephasing rates are taken as $\gamma_{21}/2\pi = 6.1~\mathrm{MHz}$, $\gamma_{31}/2\pi = 50~\mathrm{kHz}$, $\gamma_{41}/2\pi = 80~\mathrm{kHz}$, $\gamma_{51}/2\pi = 129~\mathrm{kHz}$, and $\gamma_{61}/2\pi = 6.1~\mathrm{MHz}$. The electric dipole matrix elements of the atomic transitions are, in Hartree atomic units, given by $\left| {{\mu _{12}}} \right| = 2.99$, $\left| {{\mu _{23}}} \right| = 0.00914$, $\left| {{\mu _{34}}} \right| = 211$, $\left| {{\mu _{45}}} \right| = 387$, $\left| {{\mu _{56}}} \right| = 0.0138$, and $\left| {{\mu _{61}}} \right| = 1.22$~\cite{SWM-2.5}.
These parameters define the dressed SWM dynamics and therefore determine the attainable gain, bandwidth, and dispersion of the quantum transduction response achieved by RAQR. 


%
%

We compare a conventional electronic RF receiver without atomic shaping, the single-LO SWM-based Rydberg receiver characterized by $H_\mathrm{A} \left(f\right)$ in~\eqref{eq:two_pole_lp}~\cite{chen-Rydberg3}, and the proposed atomic-comb receiver, which uses an optimized multi-tone LO to establish $H_{\mathrm{AC}}\left( f\right) $ in~\eqref{eq:H_AC_def}. The multi-tone LO is a programmable receiver-side control signal, whereas the incident OFDM RF field remains unknown and information-bearing.

We consider an OFDM system with $N=128$ subcarriers spaced at $\Delta f = 100~\mathrm{kHz}$, yielding a useful symbol duration $T_u = 1/\Delta f = 10~\mu\mathrm{s}$. The default cyclic prefix length is set to $N_{\mathrm{cp}}=16$, corresponding to a CP duration $T_{\mathrm{cp}} = \left(  N_{\mathrm{cp}}/ N \right) T_u = 1.25~\mu\mathrm{s}$. Hence, the total OFDM symbol duration is $T = T_u + T_{\mathrm{cp}} = 11.25~\mu\mathrm{s}$. Unless otherwise specified, all subcarriers are assumed active for simplicity, and the transmitted data symbols are independently drawn from QPSK or 16-QAM constellations with normalized average power $\mathbb{E}\left[ \left| X_k\right| ^2\right] =1$.
Regarding the proposed atomic-comb-based receiver, the injected LO contains $\left( 2R+1\right) $ tones with spacing $\Delta f_{\mathrm{LO}}$. Unless otherwise specified, we set $\Delta f_{\mathrm{LO}}=\Delta f$ so that the atomic-comb structure is aligned with the OFDM subcarrier (difference-frequency grid), and choose a symmetric tone set with $R=3$ or $R=4$ depending on the experiment. The complex LO weights $\left\lbrace L_r\right\rbrace _{r=-R}^{R}$ are optimized according to the ICI-only design in problem formulated in~\eqref{optimization_ICI} or the joint ICI-ISI design specified in~\eqref{optimization-joint}. 
To evaluate the difference-frequency-aware ICI suppression capability, we consider CFO and PN at the receiver. The post-CP multiplicative impairment is modeled as $	e^{\jmath \theta\left( t\right) } = e^{\jmath \left( 2\pi \Delta f_0 t + \phi\left( t\right)  \right) }$,
where $\Delta f_0$ denotes the residual CFO and $\phi\left( t\right) $ denotes the PN process. In the discrete-frequency representation, these impairments induce the Fourier coefficients $\left\lbrace C_n \right\rbrace $ in~\eqref{eq:Cn_def}.

Two channel models are considered in our simulation. The first is a flat or mildly frequency-selective channel considered for the scheme validation, so that the role of the front-end response shaping can be isolated more transparently. The second is a short tapped-delay-line multipath channel used for confirming CP/SE and tradeoff, where the effective impulse response $g \left[m\right]$ depends jointly on the propagation channel and the atomic-comb front-end. This allows us to examine how the LO-programmable atomic front-end reshapes the post-CP tail energy and thereby affects CP feasibility and spectral efficiency. Unless otherwise specified, all Monte-Carlo results are averaged over a sufficiently large number of OFDM blocks and channel/impairment realizations to ensure stable statistics.


\subsection{Atomic Comb Validation}

\begin{figure}[t]
	\centering
	\includegraphics[width=0.49\textwidth]{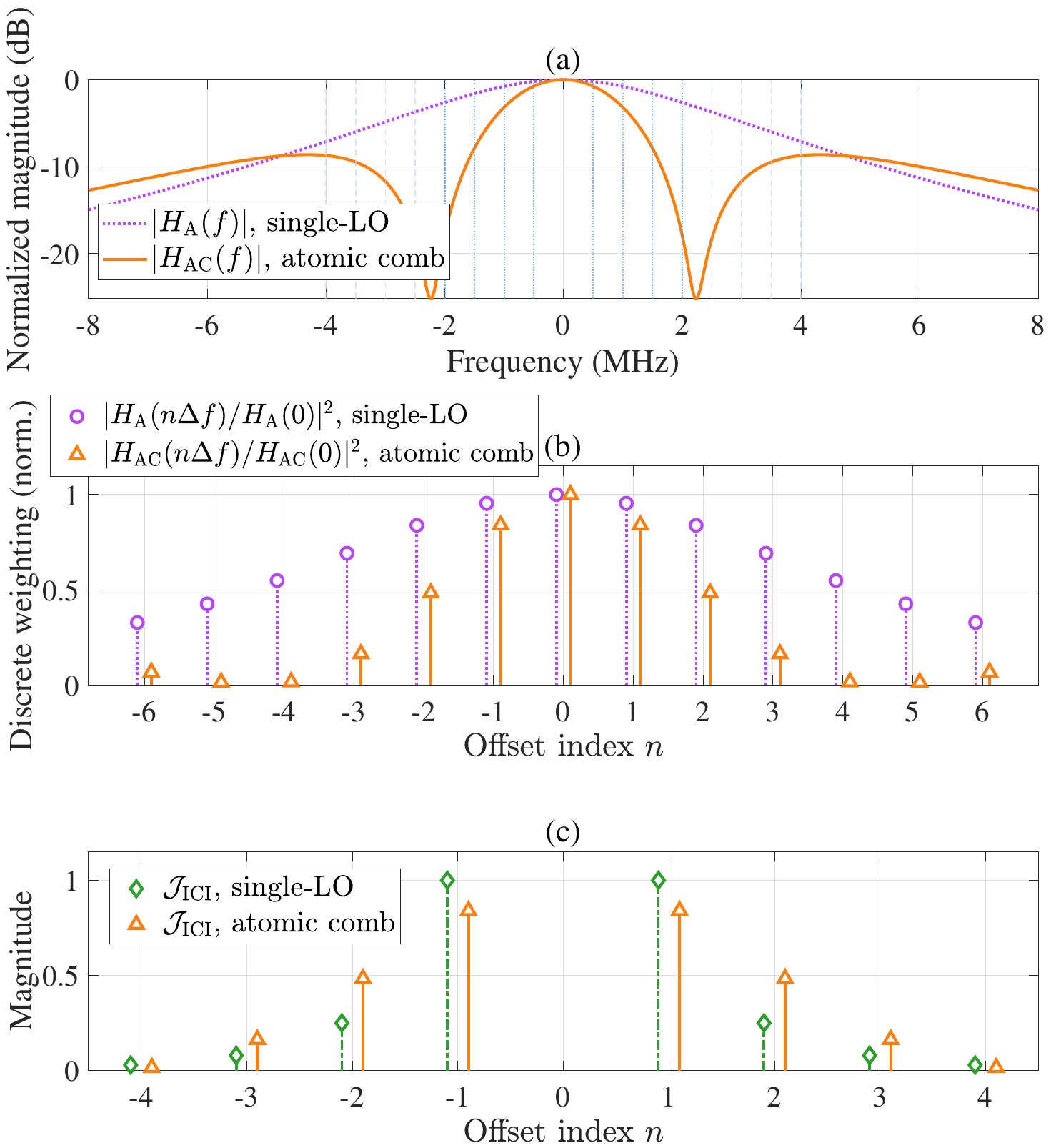}
	\caption{Performance of the proposed atomic-comb shaping. (a) Normalized effective responses, $H_\mathrm{A}\left(f\right)$ driven by a single LO and the optimized atomic-comb response $H_{\mathrm{AC}}\left(f\right)$ driven by a multi-tone LO. (b) Normalized discrete weighting sampled on the OFDM grid $f=n\Delta f$. (c) Weighted offset contributions to the aggregate ICI metric $\mathcal{J}_\mathrm{ICI}$.} \label{fig1}
\end{figure}

Fig.~\ref{fig1} validates the proposed multi-tone-LO atomic-comb shaping from three complementary perspectives. Specifically, Fig.~\ref{fig1}(a) compares the normalized effective responses, i.e.,  $H_\mathrm{A} \left( f \right) $ driven by a single LO and the optimized atomic-comb response $H_\mathrm{AC} \left( f \right) $ driven by a multi-tone LO. Unlike the smooth low-pass profile of the single-LO case, the proposed multi-tone-LO design reshapes the front-end response and generates pronounced attenuation around selected difference-frequency regions, while preserving the normalized reference response at $f=0$. 
This verifies that the injected multi-tone LO indeed provides additional degrees of freedom beyond the baseline response driven by a single LO.
Fig.~\ref{fig1}(b) further samples the two responses on the OFDM discrete grid $f=n\Delta f$. It is observed that the optimized atomic comb selectively reduces the normalized discrete weighting $\left| H_\mathrm{AC} \left( n\Delta f \right)   / H_\mathrm{AC}\left( 0\right) \right|^2$ at the dominant offset indices, especially around small $\left| n\right| $, whereas the single-LO response $\left|H_\mathrm{A} \left(n\Delta f\right)/H_\mathrm{A} \left(0\right)\right|^2$ remains comparatively large on the same grid. 
This confirms that the proposed atomic-comb shaping acts directly on the discrete difference-frequency points that determine CFO/PN-induced ICI coupling in OFDM demodulation.

Fig.~\ref{fig1}(c) further quantifies the weighted offset contributions to the aggregate ICI metric formulated in~\eqref{J_ICI}. 
For the  SWM-based RAQR driven by a single LO, the resulting weighted ICI is $\mathcal{J}_{\mathrm{ICI}}=2.47$, whereas the proposed atomic-comb design reduces it to $\mathcal{J}_{\mathrm{ICI}}=1.95$. This corresponds to a reduction ratio of $78.86\%$, i.e., an improvement of approximately $21.14\%$. It is worth emphasizing that this gain does not arise from a uniform reduction over all offset indices. Instead, the proposed multi-tone-LO atomic comb primarily suppresses the most harmful nearest-neighbor difference-frequency terms, especially around $n=\pm 1$, where the impairment weights $w_n$ are the largest. 
Although some non-dominant offsets may not be reduced monotonically, the dominant weighted contributions are effectively attenuated, thereby yielding an overall decrease in the aggregate weighted ICI metric.


\begin{figure}[t]
	\centering
	\includegraphics[width=0.49\textwidth]{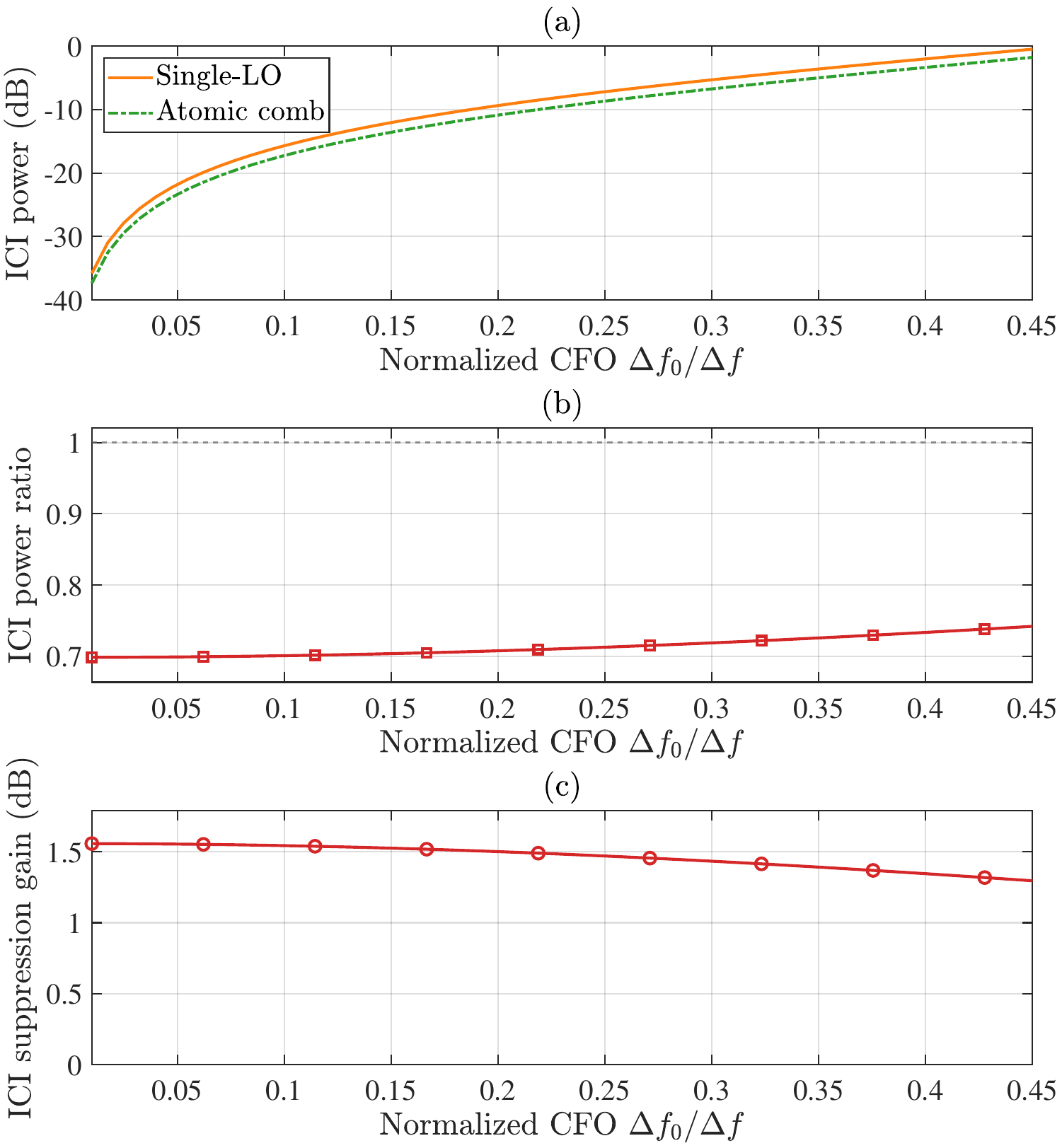}
	\caption{Post-equalization ICI suppression under CFO-only impairments. (a) Normalized post-equalization ICI power versus normalized CFO. 
		(b) ICI power ratio between the proposed atomic-comb scheme and the single-LO SWM-based RAQR. 
		(c) ICI suppression gain of the proposed atomic-comb-based receiver.} \label{fig2}
\end{figure}

We next examine whether the reduction in the weighted offset metric leads to a corresponding reduction in the post-equalization ICI power. This step is important because Fig.~\ref{fig1} primarily verifies front-end response shaping on the discrete difference-frequency grid, whereas its effect on the residual ICI power after OFDM demodulation and one-tap equalization must also be evaluated.
Specifically, we consider CFO-only impairments and denote the normalized CFO by $\Delta f_0/\Delta f$ and the CFO-induced ICI coefficient at offset order $n$ is denoted by $C_n$. 
The normalized post-equalization ICI power of the single-LO SWM-based RAQR is evaluated by 
\begin{equation}\label{P_ICI_SL}
\resizebox{\linewidth}{!}{$
\bar P_{{\text{ICI}}}^{{\text{S-L}}} = \frac{1}{{\left| {{\mathcal{K}_{\text{d}}}} \right|}}\sum\limits_{k \in {\mathcal{K}_{\text{d}}}} {\sum\limits_{n \in {\mathcal{N}_k} \cap {\mathcal{N}_{{\text{eval}}}}} {{{\left| {\frac{{{C_n}\left( {\Delta {f_0}/\Delta f} \right)}}{{{C_0}\left( {\Delta {f_0}/\Delta f} \right)}}} \right|}^2}{{\left| {\frac{{{H_{\text{A}}}\left( {\left( {k + n} \right)\Delta f} \right)}}{{{H_{\text{A}}}\left( {k\Delta f} \right)}}} \right|}^2}} } $,}
\end{equation}
where $\mathcal{N}_{\mathrm{eval}}$ denotes the evaluated nonzero ICI orders. Similarly, the normalized post-equalization ICI power of the proposed atomic-comb receiver is calculated by 
\begin{equation}\label{P_ICI_AC}
	\resizebox{\linewidth}{!}{$
\bar P_{{\text{ICI}}}^{{\text{A-C}}} = \frac{1}{{\left| {{\mathcal{K}_{\text{d}}}} \right|}}\sum\limits_{k \in {\mathcal{K}_{\text{d}}}} {\sum\limits_{n \in {\mathcal{N}_k} \cap {\mathcal{N}_{{\text{eval}}}}} {{{\left| {\frac{{{C_n}\left( {\Delta {f_0}/\Delta f} \right)}}{{{C_0}\left( {\Delta {f_0}/\Delta f} \right)}}} \right|}^2}{{\left| {\frac{{{H_{{\text{AC}}}}\left( {\left( {k + n} \right)\Delta f} \right)}}{{{H_{{\text{AC}}}}\left( {k\Delta f} \right)}}} \right|}^2}} } $} .
\end{equation}

Specifically, Fig.~\ref{fig2}(a) plots the normalized post-equalization ICI power for both schemes in dB. 
Both curves increase with the normalized CFO, since a larger CFO causes stronger leakage from neighboring subcarriers. 
Nevertheless, the proposed atomic-comb receiver consistently yields a lower post-equalization ICI power than the single-LO SWM-based baseline over the entire CFO range.
Fig.~\ref{fig2}(b) illustrates the ICI power ratio, which is defined by $\bar P_{{\text{ICI}}}^{{\text{A-C}}}/\bar P_{{\text{ICI}}}^{{\text{S-L}}}$.
Under this definition, the ICI power ratio $\bar P_{{\text{ICI}}}^{{\text{A-C}}}/\bar P_{{\text{ICI}}}^{{\text{S-L}}} < 1 $ implies that the proposed atomic-comb receiver yields lower residual ICI power than the single-LO SWM-based RAQR. Fig.~\ref{fig2}(b) shows that the ICI power ratio remains below unity for all evaluated CFO values, demonstrating that the ICI suppression is robust over the considered CFO range rather than being limited to a specific operating point. 
More explicitly, the average ratio is approximately $0.7073$, indicating that the proposed receiver eliminates the post-equalization ICI power to $70.73\%$ of the single-LO SWM-based regime. This is equivalent to an average ICI-power reduction of about $29.27\%$.
Fig.~\ref{fig2}(c) illustrates the improvement in terms of the ICI suppression gain, which is defined by $10{\log _{10}}\left( {\bar P_{{\text{ICI}}}^{{\text{S-L}}}/\bar P_{{\text{ICI}}}^{{\text{A-C}}}} \right)$.
The positive gain over the entire CFO range verifies that the proposed atomic-comb Rydberg receiver achieves consistent front-end-level ICI suppression prior to any digital post-processing. For the considered configuration, the average suppression gain is approximately $1.5$~dB.
These results are consistent with the theoretical analysis in Sec.~\ref{sec4:ici_suppression}. In particular, the post-equalization ICI power is determined by the product of the impairment-induced leakage weights $\left|C_n/C_0\right|^2$ and the receiver-side difference-frequency response weighting $\left|H\left( n\Delta f\right) /H\left( 0\right) \right|^2$. By shaping $H_{\mathrm{AC}}\left( f\right) $ over the OFDM difference-frequency grid, the proposed atomic comb suppresses the dominant weighted leakage components, thereby mitigating the aggregate post-equalization ICI power.

\subsection{Effective SE and Comb Atomic Sensitivity}
\label{subsec:nef_aware_se}

\begin{figure}[t]
	\centering
	\includegraphics[width=0.49\textwidth]{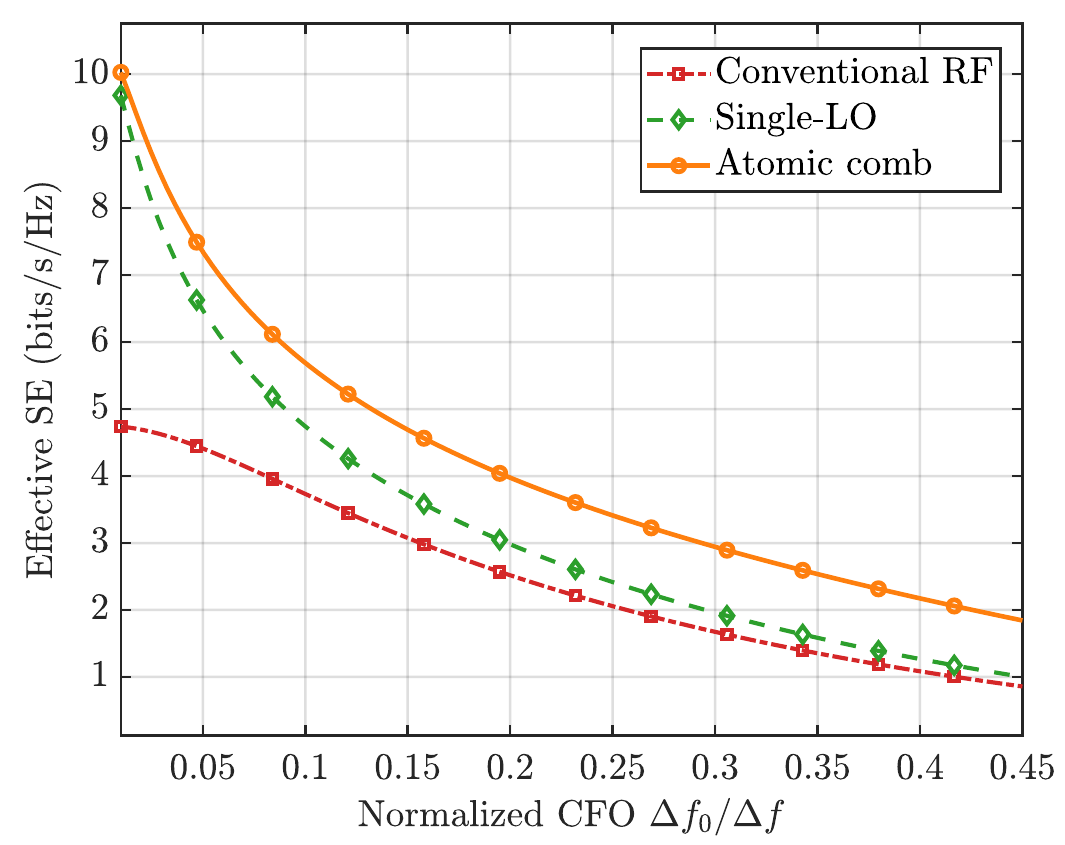}
	\caption{Effective SE versus normalized CFO $\Delta f_0 / \Delta f$.}\label{fig:nef_aware_se_cfo}
\end{figure}

Although the proposed atomic comb suppresses the post-equalization ICI under CFO-only impairments, as substantiated in Fig.~\ref{fig2}, ICI suppression alone does not fully characterize the end-to-end benefit of a practical receiver. This is because the achievable performance is jointly determined by the front-end noise floor, residual ICI, and OFDM overhead. Therefore, we further evaluate the effective SE, which translates the front-end NEF advantage and atomic-comb-based ICI suppression into a system-level rate metric.
For a given receiver scheme, the effective SINR under normalized CFO is evaluated as
\begin{equation}\label{eq:effective_sinr_nef}
{\text{SINR}}_{{\text{eff}}} = \frac{1}{{{{\left( {{\text{SNR}}_{\text{0}} } \right)}^{ - 1}} + \bar P_{{\text{ICI}}} }},
\end{equation}
where ${{\text{SNR}}_{\text{0}}^{\left( s \right)}}$ denotes the nominal receiver SNR determined by the corresponding front-end noise model, and ${\bar P_{{\text{ICI}}}^{\left( s \right)}}$ is the normalized post-equalization ICI power defined in both \eqref{P_ICI_AC} and \eqref{P_ICI_SL}. For the conventional RF receiver, ${{\text{SNR}}_{\text{0}}^{\left( s \right)}}$ is computed from the RF link budget, whereas for the SWM-based schemes it is computed from the input-referred NEF budget of the atomic receiver. The effective SE is then defined as
\begin{equation}
	\text{SE}_{\text{eff}}
	=
	\frac{ \left| \mathcal{K}_{\mathrm{d}} \right|   }{N+N_{\mathrm{cp}}}
	\log_2\left(1+\text{SINR}_{\text{eff}} \right),
\end{equation}
where $\left| \mathcal{K}_{\mathrm{d}} \right|$, $N$, and $L_{\mathrm{cp}}$ denote the number of data-bearing subcarriers, the OFDM symbol length, and the cyclic-prefix length, respectively.

Fig.~\ref{fig:nef_aware_se_cfo} shows the effective SE versus the normalized CFO. 
As observed, all schemes suffer a decrease in effective SE as the normalized CFO increases, mainly due to the stronger CFO-induced ICI. The proposed atomic comb achieves the highest effective SE over the entire CFO range, which results from the combined advantage of the SWM technique and the comb-shaped suppression of dominant difference-frequency ICI components. Compared with the single-LO SWM-based receiver, the atomic comb exhibits an additional improvement by actively shaping the front-end response over the discrete OFDM ICI grid, rather than depending only on the natural low-pass characteristic of a single-LO regime.

\subsection{Joint ICI-ISI Shaping and CP Reduction}
\subsubsection{Effective Impulse-Response Reshaping}



Although the ICI-aware atomic-comb design suppresses CFO- and phase-noise-induced leakage on the subcarrier grid, a reduced ICI metric may not necessarily imply a short effective impulse response. In particular, increased frequency selectivity may introduce a longer time-domain tail and hence increase the CP required to preserve the circular-convolution property of OFDM. We therefore next investigate whether the atomic-comb coefficients can reshape the temporal memory of the channel. We intend to reveal how the different receiver designs redistribute the effective channel energy in the delay domain and to quantify whether this redistribution translates into a smaller CP requirement under a prescribed residual-ISI constraint. 


\begin{figure}[!t]
	\centering
	\begin{subfigure}[a]{\linewidth}
		\centering
		\includegraphics[width=0.95\linewidth]{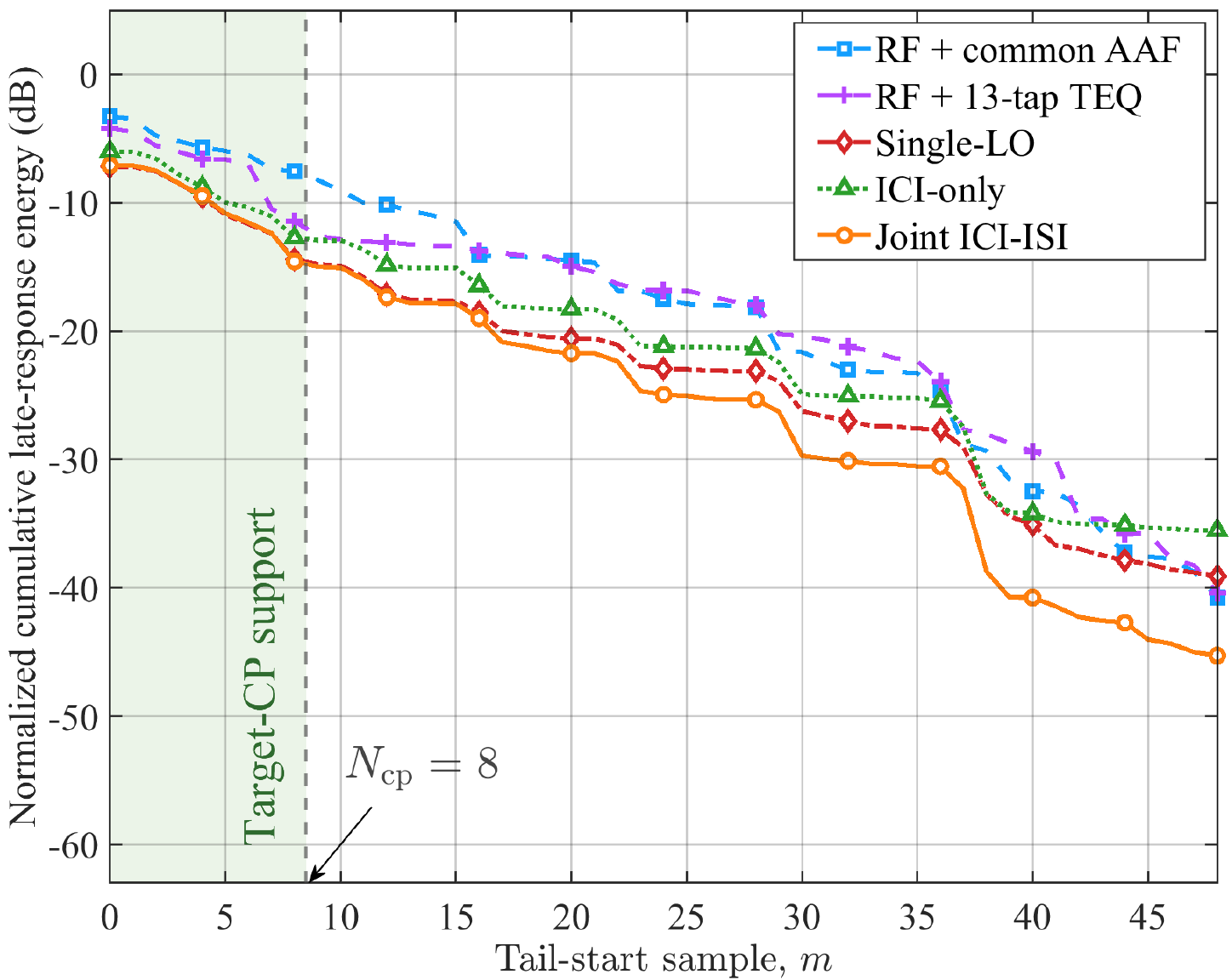}
		\caption{}
		\label{fig4_1a}
	\end{subfigure}
	\hfill
	\begin{subfigure}[b]{\linewidth}
		\centering
		\includegraphics[width=0.95\linewidth]{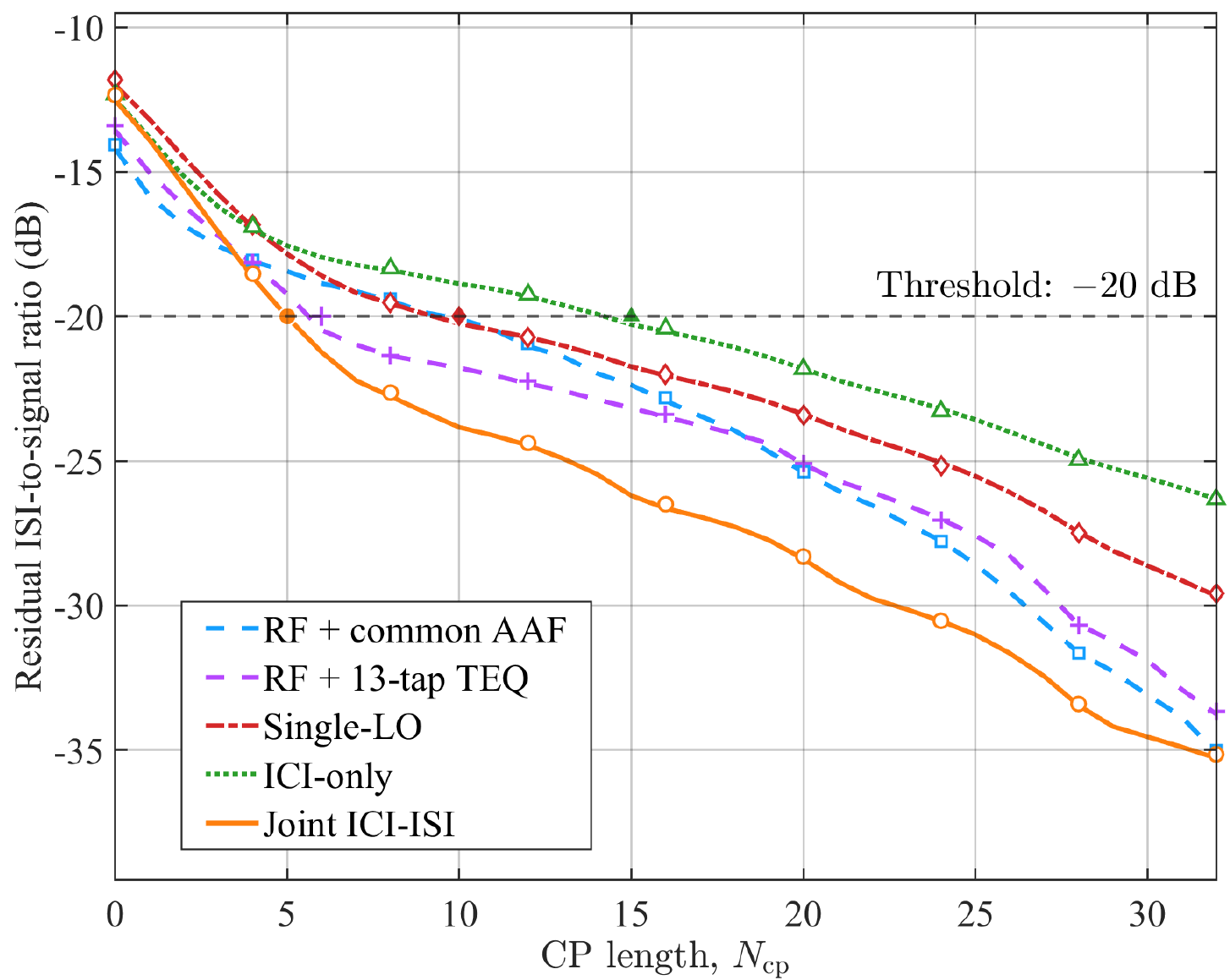}
		\caption{}
		\label{fig4_1b}
	\end{subfigure}	
	\caption{(a) Desired-signal-normalized cumulative late-response energy
		versus the candidate tail-start sample. (b) Residual ISI-to-signal ratio versus CP length.}
	\label{fig4_1}
\end{figure}


In the simulation of this section, we consider $N=128$ subcarriers, with $\left|\mathcal{K}_\mathrm{d}\right|=16$ active subcarriers and a subcarrier spacing of $\Delta f=750~\text{kHz}$, yielding a sampling rate of $96~\text{MHz}$ and an occupied bandwidth of $12~\text{MHz}$. The calibrated six-level SWM response has a 3-dB bandwidth of $10.09~\text{MHz}$. The atomic comb employs $2R+1=13$ LO tones with $R=6$ and $\Delta f_{\text{LO}}=\Delta f$, giving a comb span of $9~\text{MHz}$, comparable to the atomic bandwidth and sufficient for temporal shaping. The joint design targets $N_{\text{cp}}=8$ samples $83.33~\text{ns}$ subject to $ J_{\mathrm{ICI}}\left( \mathbf L\right)  \leq J_{\mathrm{ICI}}^{\text{S-L}}$, evaluated at $\Delta f_0 / \Delta f = 0.15$. The optimized comb weights $\mathbf L$ are then held fixed when $N_{\text{cp}}$ is swept.

To isolate the effect of receiver-side response shaping, we consider five receiver architectures, all of which are evaluated with the same wireless channel $H_{\text{ch}}\left[ k \right] $ and the same sixth-order anti-aliasing filter (AAF) $H_{\text{AAF}}\left[ k \right] $. Two additional conventional RF baselines are introduced for comparison:
\begin{itemize}
	\item RF~+~common AAF: A conventional RF receiver without either an atomic response or a digital channel-shortening filter. The term ``common AAF'' emphasizes that this filter is not unique to the RF baseline, but is also included in all atomic receiver chains. Therefore, any temporal spreading introduced by the AAF is accounted for consistently across all schemes, rather than being attributed only to the atomic front end.
	
	\item RF~+~13-tap TEQ: This baseline augments the same
	conventional RF chain with a complex-valued 13-tap time-domain equalizer
	(TEQ) before CP removal and FFT processing. The corresponding responses are
	\begin{subequations}
	\begin{align}
		G_{\text{RF}}\left[ k \right]
		&=
		H_{\text{ch}}\left[ k \right]H_{\text{AAF}}\left[ k \right],
		\label{eq:rf_common_aaf}\\
		G_{\text{TEQ}}\left[ k \right]
		&=
		H_{\text{ch}}\left[ k \right]H_{\text{AAF}}\left[ k \right]
		\sum_{q=0}^{12}w_q
		\exp\left(-\jmath\frac{2\pi kq}{N}\right).
		\label{eq:rf_teq}
	\end{align}
\end{subequations}
\end{itemize}
The TEQ length is chosen to match the 13 complex coefficients of the
13-tone atomic comb. It therefore provides a strong conventional
channel-shortening benchmark with a comparable number of design variables.
The TEQ is optimized at the same target CP length,
$N_{\text{cp}}=8$, under constraints on the passband
response, coherent gain, coefficient energy, and ICI performance.  We next examine how the different receiver architectures redistribute the
effective channel energy in the delay domain. 


Fig.~\ref{fig4_1a} portrays the normalized cumulative late-response energy that is calculated by $\sum\nolimits_{k  = m}^{N - 1} {{{\left| {g\left[k \right]} \right|}^2}} /\left[ {\frac{1}{{\left| {{\mathcal{K}_d}} \right|}}\sum\nolimits_{k \in {\mathcal{K}_\mathrm{d}}} {{{\left| {G\left[ k \right]} \right|}^2}} } \right]$, where ${g\left[ \ell  \right]}$ denotes the effective impulse response as in~\eqref{eq:g_idft} and ${G\left[ k \right]}$ represents its frequency response, respectively. For a given candidate tail-start sample $m$, this quantity measures the total effective-response energy remaining from sample $m$ onward, relative to the desired received-signal power. A lower value therefore indicates that less channel energy persists at large delays and that the effective response is more strongly concentrated near its beginning. The shaded region in Fig.~\ref{fig4_1a} indicates the delay support covered
by the target CP length $N_{\text{cp}}=8$, while the vertical
dashed line marks the corresponding CP boundary. Energy remaining beyond
this boundary is not fully absorbed by the target CP and may therefore
contribute to residual ISI. Hence, the behavior of the curves near and
beyond the dashed line directly reveals whether a receiver design is
capable of shortening the effective channel response under the prescribed
CP budget.

All curves in Fig.~\ref{fig4_1a} decrease monotonically because increasing
$m$ removes progressively more samples from the cumulative tail-energy
sum.
Around and beyond the target-CP boundary, the Joint ICI-ISI design exhibits
the lowest cumulative late-response energy, with the separation becoming
more evident at larger delays. This confirms that the joint ICI-ISI design
specifically suppresses long-delay components rather than merely reducing
the overall receiver gain. The 13-tap TEQ also shortens the effective
response, but retains more far-tail energy than the joint atomic-comb
design. By contrast, the ICI-only atomic comb exhibits a relatively strong
temporal tail because its optimization objective does not directly penalize
channel memory. In fact, additional frequency-domain selectivity introduced
for ICI suppression may broaden the corresponding time-domain response.

The cumulative late-response energy provides an intuitive view of the
delay-domain reshaping mechanism, but it is not identical to the actual
residual ISI. Residual ISI also depends on how each delayed component
overlaps the FFT observation window, the leakage from the preceding OFDM
block, and the time-domain correlation induced by activating only the
subcarriers in $\mathcal{K}_\mathrm{d}$. 
As expected in Fig.~\ref{fig4_1b}, the residual ISI decreases with $N_{\text{cp}}$ for all schemes because a longer CP accommodates a larger portion of the effective response. The rates of decrease, however, differ substantially. The joint ICI-ISI design is optimized to concentrate the effective response within the target CP rather than to minimize ISI in the absence of a CP. It is therefore not necessarily the best scheme at $N_{\text{cp}}=0$. Once the CP covers the concentrated early-delay components, its residual ISI decreases more rapidly than those of the other receivers. The single-LO SWM-based scheme remains close to the conventional RF baseline,
showing that the unoptimized atomic response alone provides little
channel-shortening benefit. The ICI-only atomic comb decreases more slowly
and eventually requires the longest CP, further indicating that
frequency-domain ICI suppression fails to automatically control
time-domain memory.

\subsubsection{ICI-ISI Pareto Tradeoff}

\begin{figure}[t]
	\centering
	\includegraphics[width=0.49\textwidth]{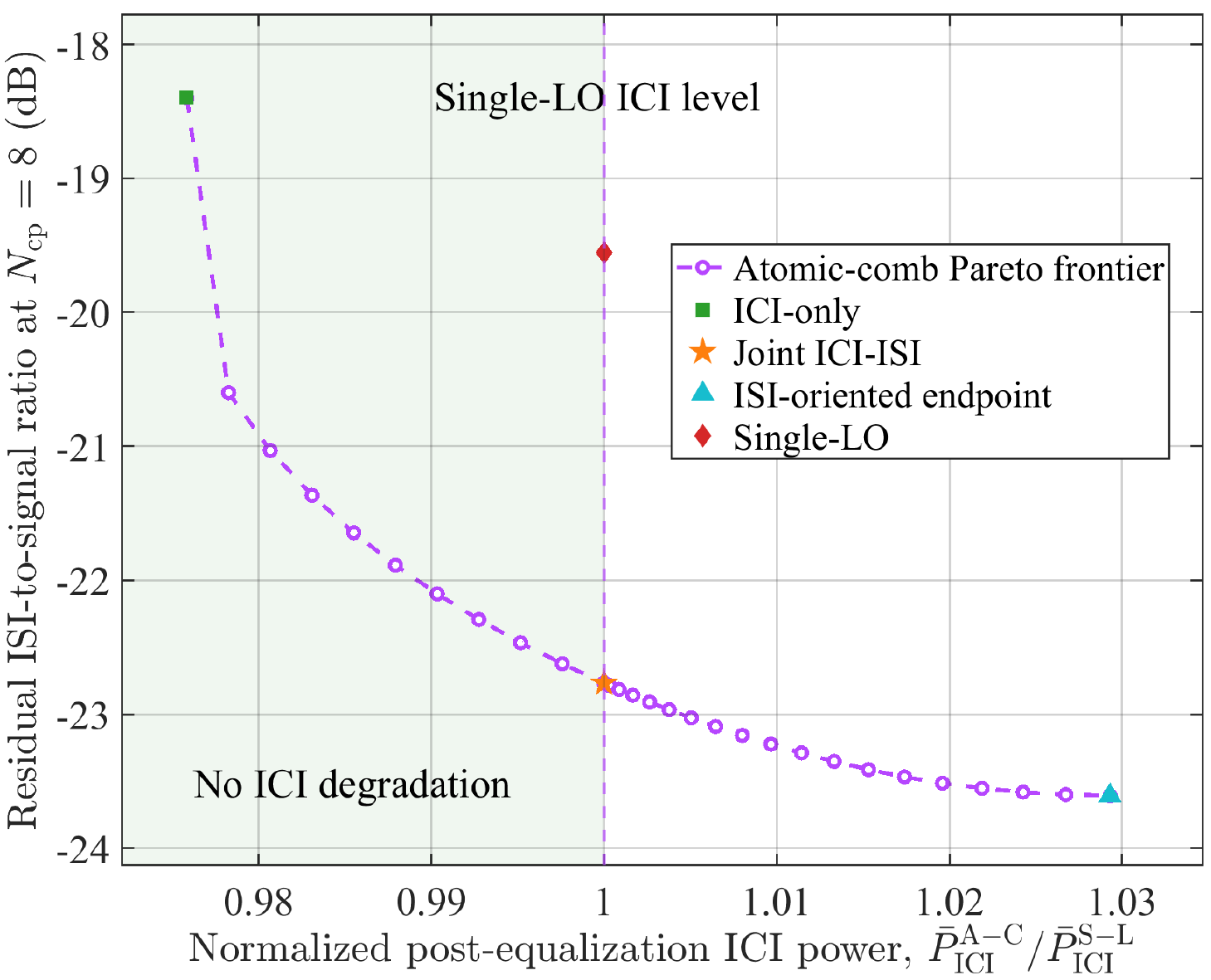}
	\caption{ICI-ISI Pareto tradeoff of the 13-tone atomic-comb
		receiver at $N_{\mathrm{cp}}=8$.} \label{fig:ici_isi_pareto}
\end{figure}

To investigate the interaction between frequency-domain ICI suppression and time-domain channel-memory shaping, Fig.~\ref{fig:ici_isi_pareto} presents the ICI-ISI Pareto frontier of the nominal 13-tone atomic-comb receiver. This simulation is motivated by the fact that minimizing ICI alone does not necessarily shorten the effective impulse response or reduce the CP requirement. By independently optimizing the atomic-comb coefficients under different allowable ICI levels, the frontier quantifies how relaxing the ICI constraint provides additional freedom to suppress delayed channel components. The monotonic decrease of the frontier confirms a fundamental tradeoff between post-equalization ICI and residual ISI, while its gradual flattening indicates diminishing ISI gains as the ICI constraint is further relaxed.

The selected joint ICI-ISI design operates at the unity boundary, where its normalized ICI power does not exceed that of the single-LO SWM-based RAQR. At the target CP length $N_{\mathrm{cp}}=8$, it reduces the residual ISI-to-signal ratio from $-19.56~\mathrm{dB}$ to $-22.77~\mathrm{dB}$, corresponding to a $3.21~\mathrm{dB}$ improvement without ICI degradation. It also outperforms the ICI-only atomic-comb design by $4.37~\mathrm{dB}$, whereas allowing approximately $3\%$ ICI degradation yields less than $1~\mathrm{dB}$ of further ISI reduction. These results place the selected design near the knee of the Pareto frontier and validate the proposed joint optimization theory: explicitly shaping temporal memory captures most of the achievable channel-shortening gain while preserving the reference ICI performance, thereby supporting practical CP reduction.

\subsubsection{Effective SE}

\begin{figure}[t]
	\centering
	\includegraphics[width=0.49\textwidth]{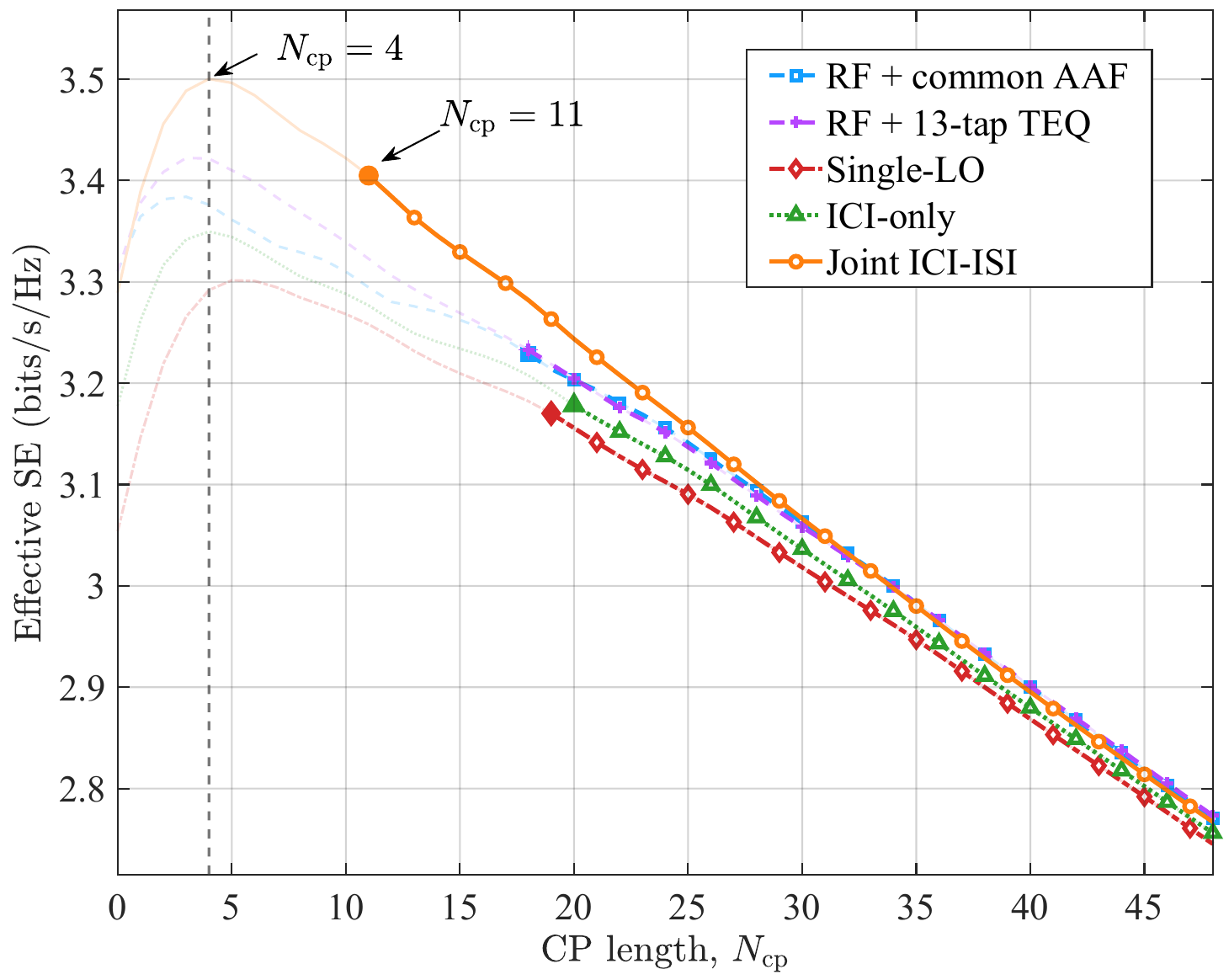}
	\caption{Effective SE versus CP length $N_{\mathrm{cp}}$.} \label{fig:effective_se_cp}
\end{figure}

To determine whether the delay-domain shortening observed above can be
translated into an end-to-end OFDM efficiency gain,
Fig.~\ref{fig:effective_se_cp} evaluates the effective SE versus the CP length $N_\mathrm{cp}$.
A common nominal post-equalization SNR of $20~\mathrm{dB}$ is adopted for all
schemes. The effective SE jointly accounts for CP overhead, exact
post-equalization ICI, covariance-aware residual ISI, and additive noise,
and is normalized by the occupied OFDM bandwidth.
To ensure a meaningful comparison, a CP length is regarded as feasible only
when the corresponding residual ISI-to-signal ratio is no greater than
$-20~\mathrm{dB}$. The pale portions of the curves represent
infeasible operating points, whereas the opaque portions satisfy the common
residual-ISI requirement. The filled marker on each curve identifies its
maximum feasible SE. 

As shown in Fig.~\ref{fig:effective_se_cp}, the joint ICI-ISI design first
satisfies the residual-ISI requirement at $N_{\mathrm{cp}}=11$, where it
achieves an effective SE of approximately $3.40~\mathrm{bits/s/Hz}$. By
comparison, the conventional RF and 13-tap TEQ baselines require
a length of approximately $N_\mathrm{cp} = 18$ to enter the feasible region, while the single-LO and ICI-only schemes require still longer CPs. The joint ICI-ISI design consequently improves the maximum feasible SE by approximately
$5.44\%$ over RF + common AAF and $5.32\%$ over the 13-tap TEQ scheme. Beyond the first feasible operating point, all curves decrease gradually since the residual ISI has already been sufficiently suppressed, whereas the CP overhead continues to increase. This confirms that the benefit of the proposed joint design is not limited to reducing a delay-domain metric. By explicitly controlling both ICI and effective channel memory, the atomic comb reduces the CP required to meet a prescribed interference level, and this reduction is directly converted into a higher SE. 

\subsubsection{SER Performance}

\begin{figure}[t]
	\centering
	\includegraphics[width=0.49\textwidth]{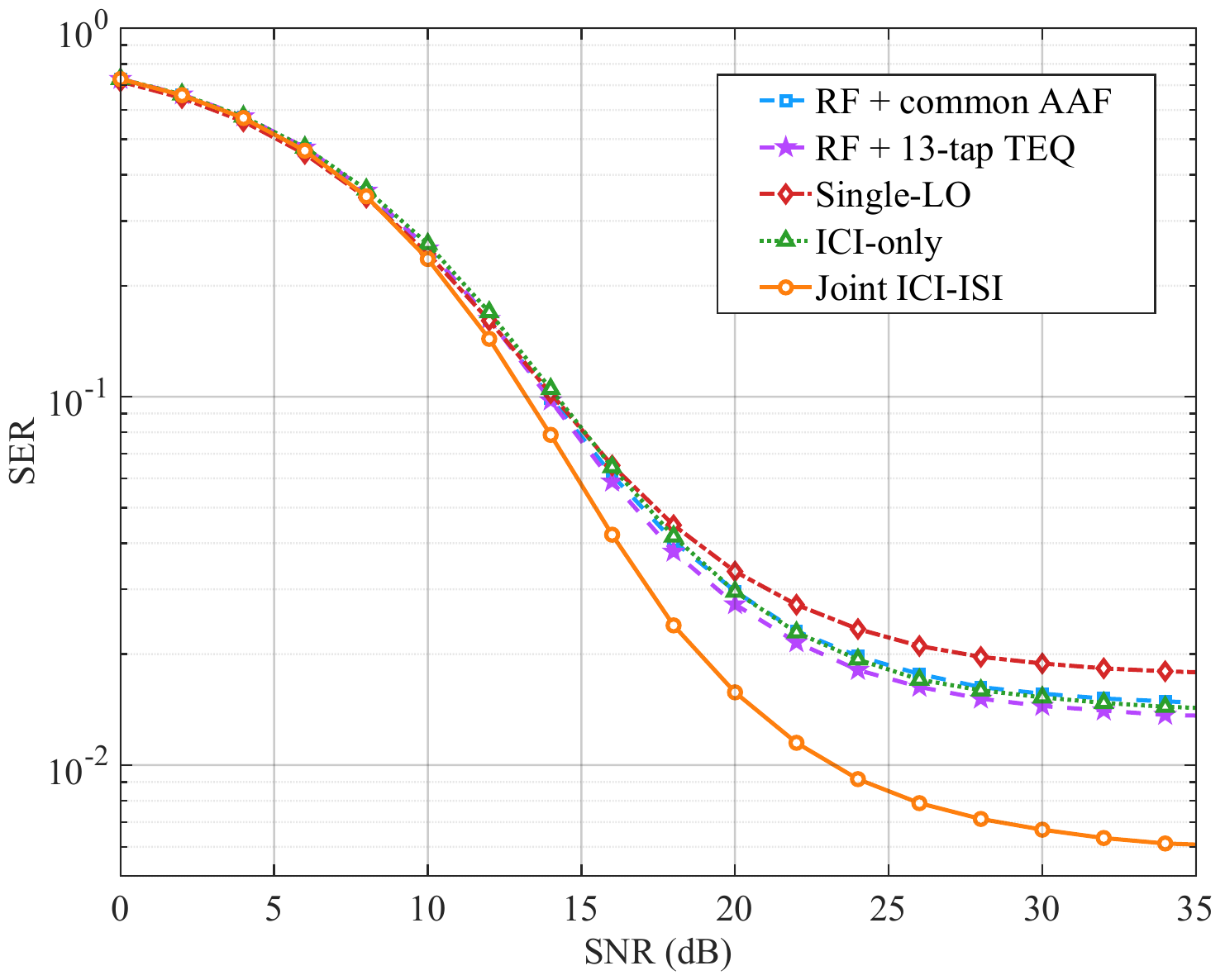}
	\caption{SER versus SNR at $N_\mathrm{cp} =11$.} \label{fig:ser_snr}
\end{figure}

Fig.~\ref{fig:ser_snr} presents SER versus SNR, in which the receiver coefficients and channel configuration are identical to
those employed in Fig.~\ref{fig:effective_se_cp}. In particular, all schemes
employ the same transmission CP length of $N_{\mathrm{cp}}=11$, which is
the shortest CP satisfying the prescribed residual-ISI requirement for
the Joint ICI-ISI receiver. The atomic-comb coefficients are designed
based on a normalized CFO of $\Delta f_0/\Delta f=0.15$, whereas the SER evaluation assumes a residual normalized CFO of $0.05$ after synchronization and CFO compensation. Furthermore, for each OFDM block, independently generated 16-ary quadrature amplitude modulation (16-QAM) symbols are mapped onto the active subcarriers and converted to the time domain using an inverse fast Fourier transform (IFFT) with the configured OFDM transform size. After CP insertion, the resulting waveform is linearly convolved with the composite impulse response associated with the considered receiver architecture. A residual CFO and additive white Gaussian noise (AWGN) are then applied to the continuous received waveform. Following CP removal, a fast Fourier transform (FFT) is performed, and the deterministic common phase rotation caused by the residual CFO is compensated. The active-subcarrier symbols are subsequently recovered using single-tap frequency-domain equalization and hard-decision 16-QAM demodulation.  All schemes employ the same transmitted symbols and noise realizations at each SNR
point for fair comparison.

As observed in Fig.~\ref{fig:ser_snr}, the five schemes exhibit similar SER in the low-SNR region because the performance is dominated by the AWGN. Their curves begin to separate at approximately $12$--$14~\mathrm{dB}$, where residual ICI and finite-CP ISI become increasingly important. The Joint ICI--ISI receiver achieves the best SER in this region. At high SNR, all schemes approach interference-limited error floors because the remaining errors are dominated by residual CFO-induced ICI and finite-CP ISI rather than thermal noise. Thus, the proposed joint design reduces the high-SNR SER by more than one half relative to the strongest conventional baselines.

\section{Conclusion}\label{sec7:comclusion}

In this paper, we have developed an atomic-comb Rydberg receiver for OFDM communications based on SWM transduction. By harnessing
a multi-tone LO, the atomic front-end response can be shaped directly on the OFDM difference-frequency grid. This capability is first exploited for CFO/PN-induced ICI suppression and was then extended to a joint ICI-ISI design by explicitly linking the frequency-domain comb response to the time-domain effective impulse response. Our simulation results demonstrate that the proposed atomic comb reduces the average post-equalization ICI power compared with the single-LO SWM-based RAQR. More importantly, the joint design confirms that ICI suppression alone does not necessarily shorten the receiver memory. By jointly controlling inter-carrier leakage and delayed response energy, the proposed design reduces the residual ISI while preserving the reference ICI level and lowers the minimum feasible CP length.

\bibliographystyle{IEEEtran}
\bibliography{Ref/ref_SWM_OFDM_trans}

\end{document}